\newcommand{\bmath}[1]{\ensuremath{\boldsymbol{#1}}}

\documentclass[useAMS,usenatbib]{mn2e}
\usepackage{amssymb}
\usepackage{graphicx}

\title[Three-dimensional shapelets]
{Three-dimensional shapelets and an automated classification scheme for dark matter 
haloes\thanks{Research undertaken as 
part of the Commonwealth Cosmology Initiative (CCI: www.thecci.org), 
an international collaboration supported by the Australian Research Council }}
\author[C.Fluke et al.]
  {C.J.~Fluke$^1$\thanks{cfluke@swin.edu.au}, A.L.~Malec$^1$, P.D.~Lasky$^{1,2,3}$,
B.R.~Barsdell$^{1}$\\
  $^1$Centre for Astrophysics \& Supercomputing, 
	Swinburne University of Technology, 
	PO Box 218, Hawthorn, Victoria, 3122, Australia\\
  $^2$Theoretical Astrophysics, Eberhard Karls University 
of T\"ubingen, T\"ubingen 72076, Germany\\
 $^3$School of Physics, University of Melbourne, Parkville, VIC 3010, Australia }

\begin{document}

\date{Accepted 18 December 2011}

\pagerange{\pageref{firstpage}--\pageref{lastpage}} \pubyear{2011}

\maketitle

\label{firstpage}

\begin{abstract}
We extend the two-dimensional Cartesian shapelet formalism to 
$d$-dimensions. Concentrating on the three-dimensional case, we derive 
shapelet-based equations for the mass, centroid, root-mean-square radius, 
and components of the quadrupole moment and moment of inertia tensors. 
Using cosmological $N$-body simulations as an application domain, 
we show that three-dimensional shapelets can be used to replicate 
the complex sub-structure of dark matter halos and demonstrate 
the basis of an automated classification scheme for halo shapes.  
We investigate the shapelet decomposition process from an algorithmic
viewpoint, and consider opportunities for accelerating the computation
of shapelet-based representations using graphics processing units (GPUs).
\end{abstract}


\section{Introduction}
Complex, three-dimensional structures abound in astronomy on all scales 
from ``fluffy'' dust aggregrates in molecular clouds (Ossenkopf 1993; 
Stepnik et al. 2003), 
to cosmological large-scale structure that has been described as 
``sponge-like'' (Gott, Dickinson \& Melott 1986), or a 
``skeleton'' (Sousbie et al. 2008) of clusters, filaments and 
voids (Barrow, Bhavsar \& Sonoda 1985; White et al. 1987).

While aspects of these structures can be expressed in terms of simple, 
geometrically-motivated properties such as their triaxiality or 
quadrupole moment,
these quantities are not able to capture the higher order complexity of 
the true shape.  The challenge, therefore, is to provide an accurate 
description of an arbitrary three-dimensional (3-d) shape, possibly over 
many physical length scales, in the hope that this can lead to 
improved theoretical or analytical insight into the structure in question.

The human visual system is more than capable of identifying structures 
and sub-structures for an individual 3-d object, but such qualitative 
interpretations only have limited use -- it is not practical to attempt 
a classification of shapes by eye when there are many thousands of 
objects to inspect.\footnote{Although, if there are enough individual 
eyes available to assist, then this approach is feasible, as the Galaxy Zoo 
project ({\tt http://www.galaxyzoo.org}) has demonstrated.}  The
preferred alternative is an automated approach including: 
\begin{itemize}
\item decomposition via an appropriate basis set 
(e.g. Fourier analysis, wavelet transformations); 
\item partitioning [e.g. Voronoi tesselation - see Icke \& van de Weygaert (1987) 
for an early cosmological application]; 
\item the use of minimal spanning trees to identify connected 
structures (Barrow et al. 1985; Pearson \& Coles 1995); 
\item Minkowski functionals [which return global geometric properties such 
as volume, surface area and edge density -- Mecke, Buchert \& Wagner (1994); 
Sahni, Sathyaprakash \& Shandarin  (1998)]; and 
\item segementation [e.g. ``dendrograms'' used by Goodman et al. (2009) to 
identify self-gravitating structures in molecular clouds].
\end{itemize}
 
The approach we present in this paper is the extension of the two-dimensional 
(2-d) shapelet method (Refregier 2003) to three dimensions.  
Shapelets are sets of orthonormal basis functions based on the Hermite 
polynomial solutions of the quantum harmonic oscillator (QHO).  
Simple analytic forms 
can be derived for the physical properties of 3-d structures (e.g. 
centre of mass, root-mean-square radius and the components of the 
quadrupole moment and moment of intertia tensors), which can be 
efficiently calculated in shapelet space.

In astronomy, 2-d shapelets have been applied to 
image simulation (Massey et al. 2004; Ferry et al. 2008), 
the morphological classification of galaxies (Kelly \& McKay 2004;
Andrae, Jahnke \& Melchior 2011) and
sunspots (Young et al.  2005), and weak gravitational lensing 
(Refregier \& Bacon 2003). The latter includes the measurement of 
shear (Kuijken 2006), flexion (Goldberg \& Bacon 2005), point-spread function 
modelling and deconvolution (Melchior et al. 2009; Paulin-Henriksson, 
Referegier \& Amara 2009), 
and weak lensing by large-scale structure from the FIRST radio 
survey (Chang, Refregier \& Helfand 2004).  Massey et al. (2007) investigated
weak lensing with polar shapelets (Massey \& Refregier 2005), a form 
more suitable for images with rotational symmetry.  Further properties 
of shapelets, including integral relations and convolution sums are presented 
in Coffey (2006).  

The importance of the shapelet approach lies not so much in the 
basis functions, but in the simplifed computation of quantities relating
to shape and structure that can be determined once a shapelet decomposition
has been obtained.  These analytic quantities are expressed as linear
sums of weighted shapelet states, greatly reducing the calculation complexity 
compared to (numerically) solving the related integral formulations.

Shapelet decomposition is not without its problems [see Berry, Hobson
\& Withington (2004) for an extensive discussion].  Melchior, Meneghetti
\& Bartelmann (2007) examined the limitations of shapelet image analysis 
in cases where the orthonormality condition [see equation (\ref{eqn:ortho}) 
below] fails, and proposed a  decomposition procedure that preserves 
physical properties of images.  Melchior et al. (2010) and Bosch (2010) 
considered problems with using circular Gaussian basis functions to 
model galaxies with high ellipticity or a large S\'{e}rsic index.
Ngan et al. (2009) proposed an alternative orthonormal basis based
on the S\'{e}rsic profile (hence S\'{e}rsiclets) for use in weak lensing
analysis.  While helping to avoid issues with poor shape recovery from
overfitting low signal-to-noise galaxies, and fitting with too many degrees
of freedom, S\'{e}rsiclets do not possess the analytic properties of shapelets, 
and the basis functions must be generated numerically.  Indeed, it is the 
existence of analytic functions that has motivated our choice of 3-d Cartesian
shapelets as an appropriate tool for quantifying properties of 
three-dimensional structures. 

The remainder of this paper is set out as follows. In Section 
\ref{sec:cartesian}, we present the mathematics of 3- and $d$-dimensional
Cartesian shapelets. New analytic expressions are presented for several 
important physical properties of 3-d structures 
in Section \ref{sct:analytic}.  In Section \ref{sct:implement}, 
we describe issues relating to implementing an efficient 3-d shapelet 
decomposition code.  We highlight the inherent high-degree of paralellism 
in the shapelet decomposition algorithm, which makes it a promising target for 
graphics processing units.
In Section \ref{sct:cosmoapp}, we present first applications of 3-d 
shapelets to problems in cosmological simulations, with an emphasis 
on studying sub-structure in dark matter halos, demonstrating how 
an automated shape classifier can work in shapelet space. We end with 
a summary and outlook for 3-d shapelets in astronomy in Section \ref{sct:conc}.

\begin{figure*}
\begin{center}
\includegraphics[width=7in]{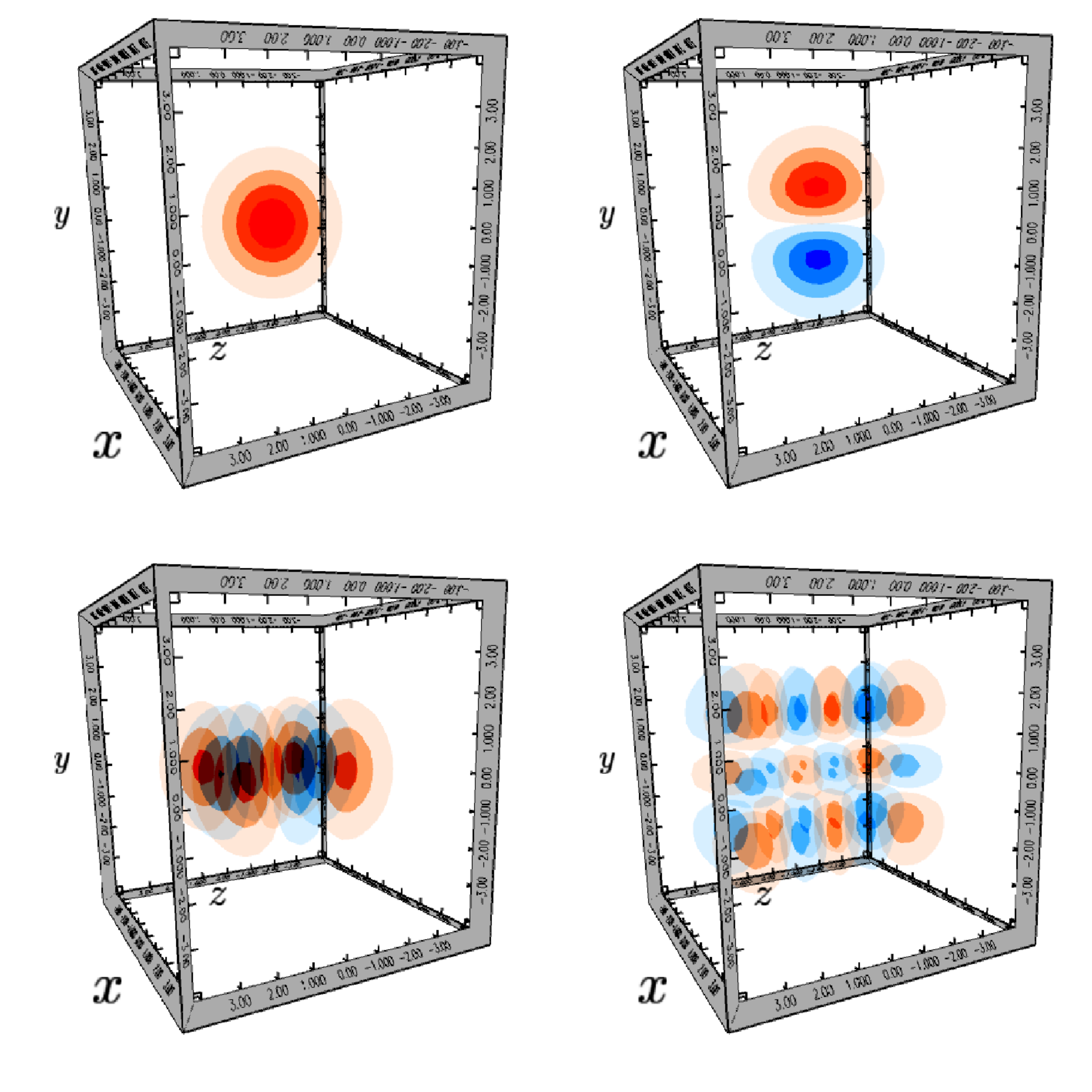}
\end{center}

\caption{Examples of three-dimensional Cartesian shapelets ($\beta = 1$).  
Top row: (left) ${\bmath n} = (0,0,0)$; (right) ${\bmath n} = (0,1,0)$.  
Bottom row: (left) ${\bmath n} = (2,0,2)$;
(right) ${\bmath n} = (1,2,4)$. For each panel, 
we calculate the maximum data value, $f_{\rm max}$, and generate 10 
equally spaced iso-surfaces over the range $(-f_{\rm max}, f_{\rm max})$.
Individual isosurfaces are coloured with a two-ended intensity colour 
map: blue $\rightarrow$ black $\rightarrow$ orange.  \label{fig:B000}}
\end{figure*}

\section{Cartesian shapelets}
\label{sec:cartesian}
In this section we present the Cartesian shapelet formalism. For full details
of the one- and two-dimensional cases, and applications, see Refregier (2003).

\subsection{One-dimensional Cartesian shapelets}
The one-dimensional (1-d) shapelet functions are 
\begin{equation}
B_n(x; \beta) \equiv \beta^{-1/2} \phi_n (\beta^{-1} x),
\label{eqn:1ddimen}
\end{equation}
where $\beta$ is a scaling length, $n$ is a non-negative integer and
\begin{equation}
\phi_{n}(x) \equiv \left(2^n \pi^{1/2} n! \right)^{-1/2} H_n(x)
e^{-x^2/2}
\label{eqn:1dshap}
\end{equation}
with $H_n(x)$ the $n$-th order Hermite polynomial.
Higher order shapelets can
be obtained using the recursion relation (see Appendix \ref{app:hermite}
for some useful expressions):
\begin{equation}
B_{n}(x;\beta) = \left(\frac{x}{\beta}\right) \sqrt{\frac{2}{n}}
B_{n-1}(x;\beta) - \sqrt{\frac{n-1}{n}} B_{n-2}(x;\beta)
\label{eqn:recur1}
\end{equation}
where
\begin{equation}
B_0(x;\beta) = \beta^{-1/2} \pi^{-1/4} e^{-x^2/2\beta^2}, 
\end{equation}
and 
\begin{equation}
B_1(x;\beta) = \frac{\sqrt{2}x}{\beta} B_0(x;\beta).
\end{equation}

The 1-d shapelets form an orthonormal basis, satisfying: 
\begin{equation}
\int_{-\infty}^{\infty} 
B_n(x; \beta) B_m(x; \beta) {\rm d}x = \delta_{nm}
\label{eqn:ortho}
\end{equation}
where $\delta_{nm}$ is the Kronecker delta symbol.  Shapelets are smooth and
continuously differentiable everywhere. 
The shapelet coeffecients for a sufficiently well-behaved 1-d function, $f(x)$,
are found through the integral:
\begin{equation}
f_n = \int_{-\infty}^{\infty} f(x) B_n(x; \beta) {\rm d}x 
\label{eqn:fn}
\end{equation}
allowing the function to be re-written as a sum of (weighted) shapelets:
\begin{equation} 
f(x) = \sum_{n=0}^{\infty} f_n B_n(x; \beta).
\end{equation}
As we show in Section \ref{sct:implement}, the calculation of $f_n$ poses the 
main computational challenge. In practice, $n$ is limited to $n \leq n_{\rm max}$ 
and the integral of equation (\ref{eqn:fn}) is calculated over a finite volume.  
However, the orthonormality condition assumes infinite support - so power 
from higher order shapelets may be lost, and the orthonormality requirement 
may no longer strictly hold if the integration region is too small 
(Melchior et al. 2007). 

\subsection{Three-dimensional Cartesian shapelets}
Using the orthonormality of 1-d shapelet functions, the 
basis functions for 2-d shapelets are (Refregier 2003): 
\begin{equation}
B_{2,\bmath{n}}(\bmath{x}; \beta) \equiv \beta^{-1} \phi_{2,\bmath{n}}
(\beta^{-1} \bmath{x}).
\label{eqn:2ddimen}
\end{equation}
where
\begin{equation}
\phi_{2,\bmath{n}}(\bmath{x}) \equiv \phi_{n_1} (x_1) \phi_{n_2} (x_2)
\end{equation}
with $\bmath{x} = (x_1, x_2)$ and $\bmath{n} = (n_1, n_2)$. 
Consequently, the extension to 3-d Cartesian shapelets is now almost trivial:  
\begin{equation}
B_{3,\bmath{n}}(\bmath{x}; \beta) \equiv \beta^{-3/2} \phi_{3,\bmath{n}} 
(\beta^{-1} \bmath{x}).
\label{eqn:3ddimen}
\end{equation}
where
\begin{equation}
\phi_{3,\bmath{n}}(\bmath{x}) \equiv \phi_{n_1} (x_1) \phi_{n_2} (x_2)
\phi_{n_3}(x_3)
\end{equation}
with $\bmath{x} = (x_1, x_2, x_3)$ and $\bmath{n} = (n_1, n_2, n_3)$. 
The 3-d Cartesian shapelet coeffecients have the form:
\begin{equation}
f_{3,\bmath{n}} = \int_V
f_3(\bmath{x}) 
B_{3,\bmath{n}}({\bmath x};\beta) {\rm d}^3 x
\label{eqn:coeff3}
\end{equation}
with the integration occuring over the infinite volume of the domain, $V$,
and the $3$-dimensional shapelet decomposition is:
\begin{equation}
f_3(\bmath{x}) = \sum_{n_1, n_2, n_3}^{\infty}
f_{3,\bmath{n}} B_{3,\bmath{n}} (\bmath{x}; \beta).
\label{eqn:decomp3}
\end{equation}
We present examples of 3-d Cartesian shapelets in Fig.~\ref{fig:B000},
using equally-spaced isosurfaces, and a two-ended intensity colour-map 
ranging from blue (negative values) to black (zero) to 
orange (positive values). 

Two further useful quantities are the characteristic radius of a 3-d shapelet: 
\begin{equation}
\theta_{3,\rm max} \approx \beta \left(n_{\rm max} + 3/2\right)^{1/2},
\label{eqn:tmax}
\end{equation}
and the size of small scale oscillatory features:
\begin{equation}
\theta_{3,\rm min} \approx \beta \left(n_{\rm max} + 3/2\right)^{-1/2}.
\label{eqn:tmin}
\end{equation}
These expressions are based on well known quantum mechanics results for the
QHO, and are the 3-d versions of the expressions presented 
in Refregier (2003).  They provide a starting point for determining 
appropriate decomposition parameters, as discussed in Section \ref{sct:minimis}.

\subsection{$d$-dimensional Cartesian shapelets}
It is straightforward to infer that the $d$-dimensional generalisation
of the shapelet basis functions is: 
\begin{equation}
B_{d,\bmath{n}}(\bmath{x}; \beta) \equiv \beta^{-d/2} \phi_{d,\bmath{n}}
(\beta^{-1} \bmath{x})
\end{equation}
with
\begin{equation}
\phi_{d,\bmath{n}}(\bmath{x}) \equiv \prod_{i=1}^{d} \phi_{n_i} (x_i) .
\end{equation}
We can then write a general orthonormality condition:
\begin{equation}
\int_V
B_{d,\bmath{n}} (\bmath{x}; \beta) B_{d,\bmath{m}} (\bmath{x}; \beta)
{\rm d}^d x
= \prod_{i=1}^{d} \delta_{n_i m_i},
\end{equation}
the shapelet coeffecients have the form:
\begin{equation}
f_{d,\bmath{n}} = \int_V f_d(\bmath{x}) 
B_{d,\bmath{n}}({\bmath x};\beta) {\rm d}^d x
\label{eqn:coeffd}
\end{equation}
and the $d$-dimensional shapelet decomposition is:
\begin{equation}
f_d(\bmath{x}) = \sum_{n_1, n_2, \dots, n_d = 0}^{\infty}
f_{d,\bmath{n}} B_{d,\bmath{n}} (\bmath{x}; \beta).
\label{eqn:coeff}
\end{equation}

In $d$-dimensions, the characteristic sizes are:
\begin{equation}
\theta_{\rm d, max} \approx  \beta \left(n_{\rm max} + d/2\right)^{1/2}
\end{equation}
and
\begin{equation} 
\theta_{\rm d, min} \approx  \beta \left(n_{\rm max} + d/2\right)^{-1/2}.
\end{equation}

We note that Coffey (2006) refers to the $d$-dimensional solutions of the
harmonic oscillator, but does not present specific $d$-dimensional 
results in the form we use.

\section{Analytic expressions}
\label{sct:analytic}
Refregier (2003) demonstrates how analytic expressions can be obtained for common
properties of 2-d images.
We now derive analytic expressions for physical properties of 3-d structures 
using 3-d Cartesian shapelets, and their generalisation to $d$-dimensions. 

\subsection{Zeroth moment}
The zeroth moment, $M_0$, of an arbitary (well-behaved) function, $f_3(\bmath{x})$, in three dimensions is 
\begin{equation}
M_0 \equiv \int_V f_3(\bmath{x}) {\rm d}^3 x.
\label{eqn:ana1}
\end{equation}
Writing this in terms of the shapelet coefficients, using equation (\ref{eqn:decomp3}),
and the orthonomality condition, equation (\ref{eqn:ortho}):
\begin{eqnarray}
\label{eqn:mass0}
M_0 &=& \sum_{n_1,n_2,n_3}^{\infty} f_{3,\bmath{n}}
\int_{-\infty}^{\infty}\!\!\!\!\!\!B_{n_1} {\rm d}x_1
\int_{-\infty}^{\infty}\!\!\!\!\!\!B_{n_2} {\rm d}x_2
\int_{-\infty}^{\infty}\!\!\!\!\!\!B_{n_3} {\rm d}x_3 \\
& = & \pi^{3/4} \beta^{3/2} \sum_{n_1,n_2,n_3}^{\rm even} f_{3,\bmath n} 
U_{3,\bmath{n}}
W_{3,\bmath{n}},
\label{eqn:mass}
\end{eqnarray}
where 
\begin{equation}
U_{n_1,n_2,n_3} \equiv 2^{(3-n_1-n_2-n_3)/2},
\end{equation}
and
\begin{equation}
W_{n_1,n_2,n_3} \equiv \left[
\left(
\begin{array}{cc}
n_1\\
n_1/2
\end{array}
\right)
\left(
\begin{array}{cc}
n_2\\
n_2/2
\end{array}
\right)
\left(
\begin{array}{cc}
n_3\\
n_3/2
\end{array}
\right)
\right]^{1/2},
\end{equation}
are factors that recur in the analytic expressions to follow.
We have used the integral property [see equation (17) of Refregier (2003)] 
for even $n$:
\begin{equation}
J_{n} \equiv \int_{-\infty}^{\infty} \!\!\!\!\!\!B_n (x; \beta)  {\rm d}x =
\left(2^{1-n} \pi^{1/2} \beta \right)^{1/2} 
\left( 
\begin{array}{c}
n \\
n/2
\end{array}
\right)^{1/2},
\end{equation}
while for odd $n$, the integrals in equation (\ref{eqn:mass0}) vanish 
as $B_n (\bmath{x}; \beta)$ is an odd function.

For applications in image processing, Refregier (2003) identifies total 
flux, $F$, with the 2-d zeroth moment.  In 3-d, a more natural association 
might be made with total mass, $M$, for an object with density field, 
$f_{3} (\bmath{x}) = \rho(\bmath{x})$. 

\subsection{Centroid}
The centroid position of a 3-d object is:
\begin{equation}
\hat{x}_i \equiv \frac{1}{M_0} \int_V x_i f_{3}(\bmath{x}) {\rm d}^3 x
\label{eqn:ana2}
\end{equation}
for $i=1,2,3$.   The orthonormality condition enables us to write the series
expansion as (for clarity, we only show results for $\hat{x}_1$): 
\begin{equation}
\hat{x}_1 = \frac{1}{M_0} \sum_{n_1,n_2,n_3}^{\infty}
f_{3,{\bmath n}} 
J_{n_2} J_{n_3}
\int_{-\infty}^{\infty} \!\!\!\!\!\!x_1 B_{n_1} {\rm d}x_1.
\end{equation}
Using the recursion relation, equation (\ref{eqn:Bn+1}),
and the fact that 
\begin{equation}
\int_{-\infty}^{\infty}\frac{ {\rm d} B_{n_{i}}}{{\rm d} x_{i}}{\rm d} x_{i}=0,
\end{equation}
gives the intermediate result
\begin{equation}
\hat{x}_1 = \frac{\sqrt{2}\beta}{M_0} \sum_{n_1,n_2,n_3}^{\infty}
f_{3,{\bmath n}} 
J_{n_2} J_{n_3}
\int_{-\infty}^{\infty} \!\!\!\! \sqrt{n_1 + 1} B_{n_1+1} {\rm d}x_1.
\end{equation}
With the notation introduced above, we have 
\begin{equation}
\hat{x}_1 = \frac{\pi^{3/4} \beta^{5/2}}{M_0}
\sum_{n_1}^{\rm odd}\sum_{n_2,n_3}^{\rm even} f_{3,\bmath{n}}
\sqrt{n_1 + 1} U_{n_1,n_2,n_3} W_{n_1 + 1, n_2, n_3}
\label{eqn:centx}
\end{equation}
and similar results for $\hat{x}_2$ and $\hat{x}_3$.

\subsection{Root-mean-square radius}
The root-mean-square (RMS) radius of a 3-d object is:
\begin{equation}
r_{\rm RMS}^2 \equiv \frac{1}{M_0} \int_V x^2 f_{3}(\bmath{x})
{\rm d}^3 x,
\label{eqn:ana3}
\end{equation}
where $x = \vert \bmath{x} \vert = \sqrt{x_1^2 + x_2^2 + x_3^2}$,
gives an estimate of the physical extent of the object under investigation.
Substituting equation (\ref{eqn:decomp3}) into the above, and 
using equation (\ref{eqn:ortho}):
\begin{eqnarray}
r_{\rm RMS}^2 & = & \frac{1}{M_0}\sum_{n_1,n_2,n_3}^{\rm even}
f_{3,{\bmath n}} 
\left[
J_{n_2} J_{n_3}
\int_{-\infty}^{\infty} \!\!\!\!\!  x_1^2 B_{n_1}{\rm d}x_1  \right. \\ 
\nonumber
& & +
\left. J_{n_1} J_{n_3}
\int_{-\infty}^{\infty} \!\!\!\!\!  x_2^2 B_{n_2}{\rm d}x_2  +
J_{n_1} J_{n_2}
\int_{-\infty}^{\infty} \!\!\!\!\!  x_3^2 B_{n_3}{\rm d}x_3 
\right] 
\end{eqnarray}
From equation (\ref{eqn:eigen}) and noting that
\begin{equation}
\int_{-\infty}^{\infty}\frac{ {\rm d^2} B_{n_{i}}}{{\rm d} x^2_{i}}{\rm d} x_{i}=0,
\end{equation}
we have
\begin{eqnarray}
\nonumber
r_{\rm RMS}^2 &= & \frac{2\pi^{3/4} \beta^{7/2}}{M_0} \sum_{n_1,n_2,n_3}^{{\rm even}}
f_{3,{\bmath n}} \left(n_1 + n_2 + n_3 + 3/2\right) \\ 
& & \times U_{n_1,n_2,n_3} W_{n_1,n_2,n_3}.
\end{eqnarray}

\subsection{Quadrupole moment tensor}
The quadrupole moment tensor is:
\begin{equation}
Q_{ij} \equiv \int_V f_{3}({\bmath x}) \left(3 x_i x_j - x^2 \delta_{ij}\right) 
{\rm d}^3 x,
\label{eqn:ana4}
\end{equation}
which is symmetric and traceless, so that there are only five 
independent elements. 
Performing the same calculations as in the previous section, the diagonal 
components of the quadrupole moment tensor are:
\begin{equation}
Q_{11} = 2\pi^{3/4}\beta^{7/2} \!\!\!\!\sum_{n_1,n_2,n_3}^{{\rm even}}
\!\!\!\!f_{3,\bmath{n}} (2n_1 -n_2-n_3)U_{n_1,n_2,n_3} W_{n_1,n_2,n_3}.
\end{equation}
$Q_{22}$ and $Q_{33}$ have a similar form.  The off-diagonal components are
\begin{eqnarray}
Q_{12} &=& 3\pi^{3/4}\beta^{7/2} \sum_{n_1,n_2}^{{\rm odd}}
\sum_{n_3}^{{\rm even}} f_{3,\bmath{n}} \sqrt{(n_1+1)(n_2+1)} \\ \nonumber
&&\times
U_{n_1,n_2,n_3} W_{n_1+1,n_2+1,n_3}
\end{eqnarray}
and similarly for the other $Q_{ij}$ with $i \neq j$.

\subsection{Moment of inertia tensor}
For the special case where $f_{3}({\bmath x})= \rho({\bmath x})$ 
represents a mass-density field, we can calculate the moments of interia.
The moment of interia tensor describes all moments of interia of an object 
about different axes of rotation, usually calculated with respect to 
the centre of mass of the object.  In component form:
\begin{equation}
I_{ij} \equiv \int_V f_{3}({\bmath x})(x^2 \delta_{ij} - x_i x_j) {\rm d}^3 x.
\label{eqn:ana5}
\end{equation}
In coeffecient space, the diagonal elements of the interia tensor are:
\begin{equation}
I_{11} = 2 \pi^{3/4} \beta^{7/2} \sum_{n_1,n_2,n_3}^{\rm even}
f_{3,\bmath{n}} (n_2 + n_3 + 1) U_{n_1,n_2,n_3} W_{n_1,n_2,n_3}
\label{eqn:ana6}
\end{equation}
and similarly for $I_{22}$ and $I_{33}$. 
The off-diagonal elements are
\begin{eqnarray}
I_{12} &=& - \pi^{3/4} \beta^{7/2} \sum_{n_1,n_2}^{\rm odd} \sum_{n_3}^{\rm even}
f_{3,\bmath{n}} \sqrt{(n_1+1)(n_2+1)} \\ \nonumber
&& \times U_{n_1,n_2,n_3} W_{n_1+1, n_2+1, n_3}
\label{eqn:ana7}
\end{eqnarray} 
and similarly for the remaining elements.

\subsection{Transformations}
Refregier (2003) demonstrates how shapelet coeffecients are modified under a general coordinate 
transformation in terms of a set of operators generating rotation, convergence, shear 
and translation.  As we have not used the operator formulation explicitly elsewhere in the present 
work, we choose not to introduce this approach now.  Instead, we treat simple coordinate 
transformations in terms of a modification of the integral in equation (\ref{eqn:coeff3}).

Consider an arbitrary (small) coordinate transformation:
\begin{equation}
\bmath{x} \rightarrow \bmath{x}' = (1 + \bmath{\Psi}) \bmath{x} + \bmath{\epsilon}
\end{equation}
where $\bmath{\epsilon}$ is a translation, and $\bmath{\Psi}$ is a $3 \times 3$
transformation matrix. To obtain the shapelet coeffecients of the 
transformed input shape, $f^T_{3,\bmath{n}}$, we must solve the integral:
\begin{equation}
f^T_{3,{\bmath{n}}} \simeq \int_V f_3(\bmath{x} - \bmath{\Psi} \bmath{x} - \bmath{\epsilon})
B_{3,\bmath{n}} (\bmath{x}; \beta) {\rm d}^3 x \\
\end{equation}
for each $\bmath{n}$, which is first order in $\bmath{\Psi}$.
We introduce transformed coordinates, and a new set of shapelet basis functions,
\begin{equation}
B_{3,\bmath{n}}(\bmath{x};\beta) \rightarrow B_{3,\bmath{n}} (\bmath{x}' - \bmath{\Psi} \bmath{x}' - \bmath{\epsilon}).
\end{equation}
In general, the relevant integral expressions for transformed coordinates must be calculated 
numerically. We can gain insight into the effect of simple transformations by considering the 
effect of translations and dilations on the shapelet ground state, 
$B_{3,\bmath{n} = (0,0,0)}(\bmath{x}; \beta)$.

\subsubsection{Translation}
The effect of a (small) translation, $\bmath{\epsilon} = \left( \epsilon_1, \epsilon_2, 
\epsilon_3 \right)$, on the shapelet coefficients is: 
\begin{equation}
f^T_{3,{\bmath{n}}} \simeq \int_{V'} f_3(\bmath{x}')
B_{3,\bmath{n}} (\bmath{x} - \bmath{\epsilon}) {\rm d}^3 x'.
\label{eqn:fTtranslation}
\end{equation}
As an example, we solve this for the $\bmath{n}$-tuples: $(0,0,0)$, $(1,0,0)$ 
and $(2,0,0)$, to find:
\begin{eqnarray}
f^T_{3,\bmath{n} = (0,0,0)} &=& 
e^{-\epsilon_1^2/4 \beta^2} e^{-\epsilon_2^2/4 \beta^2} e^{-\epsilon_3^2/4 \beta^2}\\
f^T_{3,\bmath{n} = (1,0,0)} &=&
-\frac{\epsilon_1}{\beta \sqrt{2}} e^{-\epsilon_1^2/4 \beta^2} 
e^{-\epsilon_2^2/4 \beta^2} 
e^{-\epsilon_3^2/4 \beta^2}\\
f^T_{3,\bmath{n} = (2,0,0)} &=&
\frac{\epsilon_1^2}{\beta^2 \sqrt{2}} e^{-\epsilon_1^2/4 \beta^2} 
e^{-\epsilon_2^2/4 \beta^2} 
e^{-\epsilon_3^2/4 \beta^2}.\\
\end{eqnarray}
As expected, shapelet power is transformed from the ground state to higher-order shapelet terms.
In all cases, if any of the $\epsilon_i = 0$, then the orthornormality 
condition, equation (\ref{eqn:ortho}), prevails.

\subsubsection{Dilation}
Next, we consider a transformation that is a pure dilation: 
\begin{equation}
\bmath{\Psi} = \bmath{\kappa} = 
\left(
\begin{array}{ccc}
\kappa_1  & 0 & 0 \\ 
0    & \kappa_2 & 0 \\
0    & 0 & \kappa_3 
\end{array}
\right)
\end{equation}
where all the $\vert \kappa_{i} \vert \ll 1$.
Transformed shapelet coeffecients are:
\begin{equation}
f^T_{3,{\bmath{n}}} \simeq 
\frac{\int_{V'} f_3(\bmath{x}')
B_{3,\bmath{n}} (\bmath{x}' - \bmath{\Psi}\bmath{x}') {\rm d}^3 x'}
{(1+\kappa_1)(1+\kappa_2)(1+\kappa_3)}
\label{eqn:fTdilation}
\end{equation}
Using the ground state shapelet and the same $\bmath{n}$-tuples as previously, we find:
\begin{eqnarray}
f^T_{3,\bmath{n} = (0,0,0)} &=& 
2^{3/2}
\left[2 + \kappa_1(2+\kappa_1)\right]^{-1/2} \\ \nonumber
& \times& \left[2 + \kappa_2(2+\kappa_2)\right]^{-1/2} 
\left[2 + \kappa_3(2+\kappa_3)\right]^{-1/2}  \\
f^T_{3,\bmath{n} = (1,0,0)} &=& 0 \\
f^T_{3,\bmath{n} = (2,0,0)} &=& 2 \kappa_1(2 + \kappa_1) 
\left[2 + \kappa_1(2 + \kappa_1)\right]^{-3/2} \\ \nonumber
&\times& \left[2 + \kappa_2(2+\kappa_2)\right]^{-1/2} 
\left[2 + \kappa_3(2+\kappa_3)\right]^{-1/2}  
\end{eqnarray}
Since the ground state is a symmetric shape, under a dilation, the odd shapelet 
coefficients vanish. 

\subsubsection{Rotations}
The same approach can be used for rotations about the coordinate axes,
which are defined in terms of the standard $3 \times 3$ rotation matrices 
of the form:
\begin{equation}
\mathbfss{R}_1(\theta_1) =
\left(
\begin{array}{ccc}
1 & 0 & 0 \\
0 & \cos \theta_1 & -\sin \theta_1 \\
0 & \sin \theta_1 & \cos \theta_1
\end{array}
\right),
\end{equation}
and similarly for rotations about the $x_2$-axis, 
${\textbfss R_2}(\theta_2)$, and $x_3$-axis, ${\textbfss R_3}(\theta_3)$. 
A sequence of rotations can be combined into a single general rotation 
matrix, ${\textbfss R}_{\bmath{x}}(\bmath{\theta})$.
The coordinate transformations for rotations are tractable but more 
complex algebraically than for translations and dilations -- equations
(\ref{eqn:fTtranslation}) and (\ref{eqn:fTdilation}).  
Rather than providing a general analytic 
form for the rotations, we instead demonstrate the resulting change 
in amplitude of shapelet coeffecients under an abitrary rotation in 
Section \ref{sct:auto}, in particular Figs.~\ref{fig:rot1}-\ref{fig:rot3}.

\subsection{$d$-dimensional expressions}
We can use the results from the previous sub-sections to obtain 
analytic expressions in $d$-dimensions. 
The zeroth moment is:
\begin{equation}
M_0 = \pi^{d/4} \beta^{d/2} \sum_{n_1,n_2, \dots, n_d}^{{\rm even}}
f_{d,{\bmath n}}
U_{d,{\bmath n}}
W_{d,{\bmath n}}
\end{equation}
where now
\begin{equation}
U_{n_1, n_2, \dots, n_d} = 2^{\frac{1}{2}\left(d - \sum_{i=1}^{d} n_i \right)}
\end{equation}
and
\begin{equation}
W_{n_1,n_2, \dots, n_d} =
\left[ \prod_{i=1}^d 
\left(
\begin{array}{cc}
n_i \\
n_i/2 \\
\end{array}
\right)
\right] ^{1/2}.
\end{equation}
The centroid is:
\begin{eqnarray}
\hat{x}_1 &=& \frac{\pi^{d/4} \beta^{(d+2)/2}}{M_0}
\sum_{n_1}^{{\rm odd}}
\sum_{n_2, \dots, n_d}^{{\rm even}}
f_{d,{\bmath n}}
\sqrt{\left(n_1+1 \right)} \\ \nonumber
&& \times U_{n_1,n_2, \dots, n_d}
W_{n_1+1, n_2, \dots, n_d},
\end{eqnarray}
and similarly for $\hat{x}_2, \dots, \hat{x}_d$.  
Finally, with $x = \vert {\bmath x} \vert = \sqrt{x_1^2 + x_2^2 + \dots +
x_d^2}$, we have the $d$-dimensional RMS radius:
\begin{eqnarray}
r_{\rm RMS}^2\!\!\!&=& \!\!\!\frac{2 \pi^{d/4} \beta^{2 + d/2}}{M_0}
 U_{n_1, \dots, n_d} W_{n_1, \dots, n_d}  \\ \nonumber
&&\times \sum_{n_1, \dots, n_d}^{\rm even} f_{d,{\bmath n}} \left(n_1 + n_2 + \dots + n_d + 
\frac{d}{2} \right). 
\end{eqnarray}

We do not attempt to derive $d$-dimensional equivalents of the 
quadrupole moment or moment of inertia tensors, as these are more natural
quantities in three-dimensions.   However, the generalised approach we 
have demonstrated can be applied to other properties defined 
as $d$-dimensional integrals of $f_{d}({\bmath x})$.

\section{Implementation Issues}
\label{sct:implement}
Before we can use the analytic expressions of Section \ref{sct:analytic} 
to study three-dimensional objects, we need to obtain the shapelet 
coefficients.  In this section, we discuss some of the issues in implementing  
an effecient 3-d shapelet decomposition code.  

\subsection{Voxellation}
\label{sct:voxel}
In applications to image simulation (Massey et al. 2004; 
Young et al. 2005) and gravitational lensing (Refregier \& Bacon 2003; 
Goldberg \& Bacon 2005; Kuijken 2006), shapelet quantities are calculated 
for a pixel grid of image intensities, which is often obtained as 
a `postage stamp' region selected from a larger image. For the 3-d case, 
we use a regular cubic mesh of voxels (volume elements). 

The discrete sampling of the 3-d structure onto a mesh means we need 
to integrate each shapelet term over the physical size of a voxel, under
the assumption that the data value in the voxel is constant.  This is
valid for data that is already on a grid (e.g. from a mesh-based simulation), 
and can be achieved for point-based data by smoothing
to the grid with an appropriate smoothing scheme.  

For integration over a finite cubic volume, $\hat{V}$, 
over spatial range $x_{\rm min}$ to $x_{\rm max}$ (and similarly
for $y$ and $z$), equation (\ref{eqn:coeff3})
is replaced by a summation over $N_g^3$ voxels: 
\begin{equation}
f_{3, \bmath{n}}  = \sum_{i,j,k}^{N_g,N_g,N_g} f_{ijk} \int_{\hat{V}_{ijk}} 
B_{\bmath n}({\bmath x}) {\rm d}^3 x
\label{eqn:bigsum}
\end{equation}
where our grid-based 3-d shape has a constant value in each voxel, $f_{ijk}$.
The volume is assumed to be sufficiently large that the $f_{ijk} \rightarrow 0$
outside of the integration region.

Following Massey \& Refregier (2005), the orthonormality of shapelets means
we can simplify the per-voxel integration of the shapelet term as the 
product of three one-dimensional integrals of the form:
\begin{equation}
I_{n}(i) = \int_a^b B_n (x) {\rm d}x.
\label{eqn:integral}
\end{equation}
where the index, $1 \leq i \leq N_g$, specifies the one-dimensional 
voxel coordinate, and hence the integration limits on the boundaries
of the $i$th voxel are: 
\begin{eqnarray}
a & = & x_{\rm min} + (i-1) \Delta x \\
b & = & a + \Delta x,
\end{eqnarray}
with cell width
\begin{equation}
\Delta x  =  \frac{x_{\rm max} - x_{\rm min}}{N_g}.
\end{equation}
This allows us to write equation (\ref{eqn:bigsum}) as a sum over all voxels:
\begin{equation}
f_{3,{\bmath n}} 
= \sum_{i,j,k}^{N_g,N_g,N_g} f_{ijk} I_{n_1}(i) I_{n_2}(j) I_{n_3}(k)
\label{eqn:bigsum2}
\end{equation}
providing a set of shapelet coeffecients that are used to calculate
the analytic quantities of Section \ref{sct:analytic}.   

Equation (\ref{eqn:integral}) has recursion 
solutions\footnote{There is an error in the factors of $\beta$ in  equation (32) 
of Massey \& Refregier (2005), which is corrected in the arXiv version of their paper: 
arXiv:astro-ph/0408445.}
\begin{equation}
I_{n}(i) = -\beta \sqrt{\frac{2}{n}} \left[ B_{n-1}(x) \right]_a^b
+ \sqrt{\frac{n-1}{n}} I_{n-2}(i)
\label{eqn:rec1}
\end{equation}
with
\begin{eqnarray}
\label{eqn:I0}
I_0(i) & = & \sqrt{\frac{\beta \pi^{1/2}}{2}} \left[ {\rm erf}
\left(\frac{x}{\beta \sqrt{2}}\right)
\right]_a^b \\
I_1(i) & = & -\beta \sqrt{2} \left[ B_0(x) \right]_a^b.
\end{eqnarray}

\subsection{Optimal decomposition}
\label{sct:minimis}
A key problem is the choice of parameters, 
($\beta$, $n_{\rm max}$, ${\bmath x}_c$),  to 
perform an optimal shapelet decomposition. We use
the notation ${\bmath x}_c$ to refer to the best-fitting object centroid, 
as opposed to the shapelet reconstructed value, $\hat{\bmath x}$.  
A good choice of parameters will ensure 
compact representation of the original data in coefficient space,
while retaining high accuracy. Well chosen 
parameters will also exclude any noise that may be present in the data. 
As Melchior et al. (2007) highlighted for the 2-d case, shapelet decompositions 
may appear good visually, so it is important to define an appropriate goodness 
of fit, particularly as shapelet space can be highly degenerate. 

The $\beta$ parameter is the characteristic scale of the object to 
be decomposed.  Increasing $\beta$ has the effect of increasing the 
amplitude of the shapelets and dilating them along all coordinate axes. 
Changing the amplitude of the shapelets has no effect on the 
optimisation as the obtained coefficients 
simply scale in proportion to the change in amplitude, i.e. $\beta$ 
is a one-dimensional spatial parameter.

The maximum number of coefficients needed relates to the complexity of 
the data. A value of $n_{\rm max}$ that is too low will likely result in loss 
of information regarding the smallest features; if $n_{\rm max}$ is too 
high, noise and arbitrary high-frequency variations will be reproduced. 
Moreover, with increasing $n_{\rm max}$, the range of $\beta$ and 
${\bmath x}_c$ values that give viable solutions increases.  This is 
because the additional coefficients can 
compensate for a poor choice of $\beta$ and ${\bmath x}_c$.
It is therefore important that a minimum optimal $n_{\rm max}$ 
value is used, while not resulting in significant loss of structural 
information, along with the optimal $\beta$ and ${\bmath x}_c$ values.

To determine appropriate $n_{\rm max}$ and $\beta$ values, we 
solve for the two unknown quantities in equations (\ref{eqn:tmax}) and 
(\ref{eqn:tmin}):
\begin{equation}
n_{\rm max} = \frac{\theta_{3, {\rm max}}}{\theta_{3, {\rm min}}} - \frac{3}{2} 
\end{equation}
and 
\begin{equation}
\beta = \sqrt{\theta_{3, {\rm max}} \,\, \theta_{3,{\rm min}}}.
\end{equation}

Consider a voxel grid centred on the coordinate origin with major axis 
length, $x_{\rm max} = -x_{\rm min}$, which is taken to be twice the 
maximum particle distance, $\theta_{3,{\rm max}}$, from the coordinate 
origin. In this case, the cell width is:
\begin{equation}
\Delta x = \frac{2 x_{\rm max}}{N_g}.
\label{eqn:deltax}
\end{equation}
Choosing $\theta_{3,{\rm min}} = \Delta x/2$, it follows that 
\begin{equation}
n_{\rm max} = (N_g-3)/2
\end{equation}
and 
\begin{equation}
\beta = x_{\rm max}/\sqrt{2 N_g}.
\label{eqn:betaval}
\end{equation}
For specific applications, convergence studies may be a more appropriate
way to select initial estimates for $n_{\rm max}$ and $\beta$, and
the size of the data `padding' region.

To minimise the number of evaluations, and avoid some of the issues
of generating shapelet coeffecients with too many orders, we impose 
the constraint [see Section 3.1 of Refregier (2003)]:
\begin{equation}
0 \leq \left( n_1 + n_2 + n_3\right) \leq n_{\rm max}.
\label{eqn:constraint}
\end{equation}
This constraint means that the total number of shapelet
terms to be evaluted for a given $n_{\rm max}$ is:
\begin{equation}
N_{\rm eval} = \frac{1}{6} (n_{\rm max}+1)(n_{\rm max}+2)(n_{\rm max}+3).
\end{equation}
This last equation is the $d=3$ version of the more 
general result: 
\begin{equation}
N_{\rm eval} = \left(
\begin{array}{c}
n_{\rm max} + d\\
d
\end{array}
\right)
\end{equation}
to obtain the unique set of ${\bmath n}$ values satisfying: 
\begin{equation}
0 \leq \sum_{m}^{d} n_m \leq n_{\rm max}. 
\end{equation}

Coefficient-based measurements may produce inaccurate results in cases 
when the chosen $n_{\rm max}$, $\beta$ and ${\bmath x}_c$ values result 
in a reconstructed shape that is truncated by the bounding cube of the 
original data grid (in other words, when the reconstructed shape 
is bigger than the original data). Ensuring that the parameter bounds 
previously outlined are not traversed, i.e. through the use of the
padding region,  will prevent this from occurring.  
Moreover,
estimates of the $\hat{{\bmath x}}$ and $r_{\rm RMS}^2$ may fail if
$M_0 = 0$, since they depend on the reciprocal of the zeroth moment.  
This may occur for values of $\beta$ that are too large.

Further discussion of strategies for optimal shapelet decomposition are 
beyond the scope of this paper -- see Massey \&
Refregier (2005) for an approach based on the steepest descent method.
We now investigate
the decomposition process from an algorithmic viewpoint, and consider 
opportunities for accelerating the computation of shapelet coeffecients 
using graphics processing units.

\subsection{Algorithmic considerations}
The algorithm for obtaining a shapelet decomposition for a voxellated 
structure is: 
\begin{enumerate}
\item Choose the target grid resolution, $N_g$, and desired $n_{\rm max}$,
which constrain the initial choice of $\beta$. 
\item Generate an array of shapelet amplitude estimates, ${f}_{3,\bmath{n}}$, 
with $N_{\rm eval}$ entries (i.e. the minimum number that must be calculated),
and initialise to zero-values. 
\item Calculate the 1-dimensional $I_n(i)$ terms for all 
orders up to $n_{\rm max}$, resulting in $n_{\rm max} N_g$ stored values 
of $I_n(i)$. 
\item Loop over the elements of the ${\bmath n}$ vector, subject to the constraint 
of equation (\ref{eqn:constraint}), then:
\begin{enumerate}
\item For each set of ${\bmath n}$ values, loop over $N_g^3$ cells with 
indices $(i,j,k)$ 
and calculate the quantity:
\begin{equation}
f_{3,\bmath{n}} := f_{3,\bmath{n}} + I_{n_1}(i) \times I_{n_2}(j) 
\times I_{n_3}(k) \times f_{ijk}.
\label{eqn:looper}
\end{equation}
\end{enumerate}
\item Output the shapelet amplitudes for further processing and analysis. 
\end{enumerate}

The process for reconstructing a three-dimensional shape from its shapelet 
coeffecients proceeds as follows:
\begin{enumerate}
\item Create an empty shape, $\hat{f}_3(\bmath{x})$, 
with dimensions $N_g^3$, and zero all $\hat{f}_{ijk}$ values.
\item Loop over ${\bmath n}$ vector, subject to constraint of equation 
(\ref{eqn:constraint}), calculating:
\begin{equation}
\hat{f}_3(\bmath{x}) := \hat{f}_3(\bmath{x}) + B_{3,\bmath{n}}(\bmath{x};\beta).
\end{equation}
\end{enumerate}

An optional filter can be applied in the reconstruction by only adding the contributions
from shapelet terms where $f_{3,\bmath{n}}$ meets a prescribed criteria.  Such 
an approach may be useful for removing noise, or to investigate the dependence of the
analytic solutions on a particular shapelet order --  see the example
application in Section \ref{sct:quant}.

Obtaining a shapelet decomposition of a voxellated structure involves
computing equation (\ref{eqn:bigsum2}) for all $N_{\rm eval}$ coefficients. This
computation is both very regular and abundant in inherent parallelism
-- two traits that suggest a strong suitability for implementation on
many-core computing architectures such as graphics processing units
(GPUs).

GPUs were originally developed to accelerate the rendering of
three-dimensional graphics through the use of a custom processor with
a highly parallel architecture. GPUs are now capable of supporting
general (i.e. non-graphics) computations through the use of software
platforms such as the Compute Unified Device Architecture (CUDA) from
NVIDIA\footnote{http://www.nvidia.com/object/cuda\_home\_new.html} or
implementations of the OpenCL\footnote{http://www.khronos.org/opencl/}
standard.
  
We can assess the suitability of the shapelet algorithm for a GPU
implementation by using an algorithm analysis approach similar to that
of Barsdell, Barnes \& Fluke (2010), who noted that the most important
considerations for an algorithm on a GPU are: massive parallelism,
branching, arithmetic intensity and memory access patterns. To begin
with, we assess the amount of parallelism in the shapelet decomposition
problem.  For simplicity, we assume that data is placed inside a bounding 
box such that all of the dimensions are the same -- if there are fewer 
grid points along one axis, these must be zero-padded to the maximum grid scale.

The computation of equation (\ref{eqn:bigsum2}) involves a summation over the
three coordinate dimensions, $i, j, k$, for each shapelet coefficient
defined by $n_1, n_2, n_3$. The computation over $n_1, n_2, n_3$ is
therefore entirely (or \textit{embarassingly}) parallel, as each
coefficient can be computed independently. The summation over voxels
also exhibits inherent parallelism, but requires some coordination
between elements. For this reason we will first consider parallelising
the shapelet algorithm only over the shapelet coefficients, and will
assume the summations are performed sequentially.
  
Parallelising the problem over the shapelet coefficients defined by
$n_1, n_2, n_3$ allows a maximum of $N_{\rm eval}$ parallel
\textit{threads} to work on the problem simultaneously. For
$n_{\rm max} = 20$, this is 1771 threads. While this is likely to
exceed the number of physical processor cores in any current hardware
architecture, modern GPUs often require an order of magnitude more
threads in flight before their full potential is reached. It is
therefore likely that some of the summation will need to be
parallelised in addition to the evaluation of the shapelet
coefficients. One such approach would be to compute sums over slices of
the data volume in parallel, before combining them in a second stage
of computation. This would increase the number of threads by a factor
of $N_g$, which would almost certainly saturate the available hardware
performance.
  
The next concern is branching, which occurs when parallel threads
execute differing instructions as a result of a conditional statement.
Besides the application of the constraint
$n_1 + n_2 + n_3 < n_{\rm max}$, the shapelet decomposition algorithm
does not require any branching operations. We therefore conclude that
this factor will not significantly influence performance on a GPU.
  
Arithmetic intensity is the ratio of arithmetic operations to
memory-access operations. A high arithmetic intensity means that the
GPU's instruction hardware will be fully utilised; a low intensity
means that getting data from memory to the processors will be a
bottle-neck and performance will be limited. The total input data to
the shapelet algorithm scales as $O(N_g^3)$, while the computation
scales as $O(N_{\rm eval} N_g^3)$. This implies a very high theoretical
peak arithmetic intensity of $O(N_{\rm eval}) \approx O(1000)$. This would
be achieved by re-using input data $f_{ijk}$ for the computation of
many $n_1, n_2, n_3$ values. Assuming such behaviour could be effected,
the performance would be limited by the arithmetic throughput of the
hardware, and we would expect to see very good performance on a GPU.
  
In practice, the re-use of data is achieved through the exploitation of
a {\em cache}, which is an area of very fast memory in which small
amounts of data can be stored. On NVIDIA GPUs, the specific cache we
refer to is known as {\em shared memory}. By loading a block of 
$f_{ijk}$ data into shared memory, threads can re-use the data
multiple times before having to load another block in. As all of the
operations on the input data $f_{ijk}$ scale as $O(N_g^3)$, there is
no difference between cacheing a block of any particular shape; for
simplicity we therefore consider cacheing a simple one-dimensional
block of data in the $i$ dimension. In this setup, the $j$ and $k$
indices can remain constant during the computation of the block, which
allows the value $I_{n_2}(j) I_{n_3}(k)$
to be pre-computed and stored locally before computation
of the block begins. If, in addition, the value of $n_1$ is made to
remain constant over the local group of threads, then the values
$f_{ijk}I_{n_1}(i)$ can be pre-computed and stored in shared
memory. The computation by each thread of the block of $i$ values then
only involves the multiplication and accumulation of two pre-computed
values. Multiplication followed by addition also happens to be the
fastest operation available on current GPU hardware.
  
The last concern is the memory access pattern exhibited by the
algorithm. Fortunately, the regularity of the computation means that
data are typically accessed in an aligned and contiguous fashion, and
there should therefore be no issues in achieving a high memory
throughput.
  
To reduce the computational overhead in the evaluation of equation 
(\ref{eqn:bigsum2}), the integrals $I_n$ can be pre-computed once for each 
input shape and stored in look-up tables.  The recursion relation, 
equation (\ref{eqn:rec1}), makes it practical to evaluate
and store {\em all} shapelet orders up to $n_{\rm max}$ for $N_g$
grid cells along one dimension. This involves only 
$O(n_{\rm max} N_g)$ terms, and could be computed on the CPU without
impacting on the overall performance of the algorithm. A further
advantage of using the recursion relations is that sufficient 
numerical precision can be maintained, even for high $n$ values.  If we 
calculated each shapelet term independently from equation (\ref{eqn:1dshap}), 
then the pre-factors, $\left( 2^n n! \right)^{-1/2}$ tend to zero very rapidly, 
and for $n>30$, cannot be stored sufficiently accurately in single 
precision.  This requirement is of relevance to GPU implementation, 
as the greatest processing speed-ups offered by the current generation 
of GPUs is for single precision, and reduces the 
overall memory required by storing as 32-bit rather than 64-bit values.

Given the strong degree of parallelism exhibited by the algorithm, the
ability to efficiently cache the input data and take
advantage of a very high arithmetic intensity, the ability to
pre-compute the shapelet integral terms, and the fact that the core of
the algorithm can be reduced to simple multiply-add operations, we
conclude that an implementation of the shapelet decomposition algorithm
on a GPU would likely achieve a level of performance very near the peak
capability of the hardware. Shapelet decomposition thus stands to
benefit significantly from current trends in commodity computing
hardware, and may have an additional advantage over related methods that
are unable to take advantage of massively-parallel architectures.
  
The extension of the above algorithm analysis to $d$-dimensional shapelet
decompositions should be straightforward, and we expect the conclusions
to remain unchanged; however, implementation complexity is likely to
increase, particularly in the general case.

\begin{figure}
\begin{center}
\includegraphics[width=3.4in,angle=0]{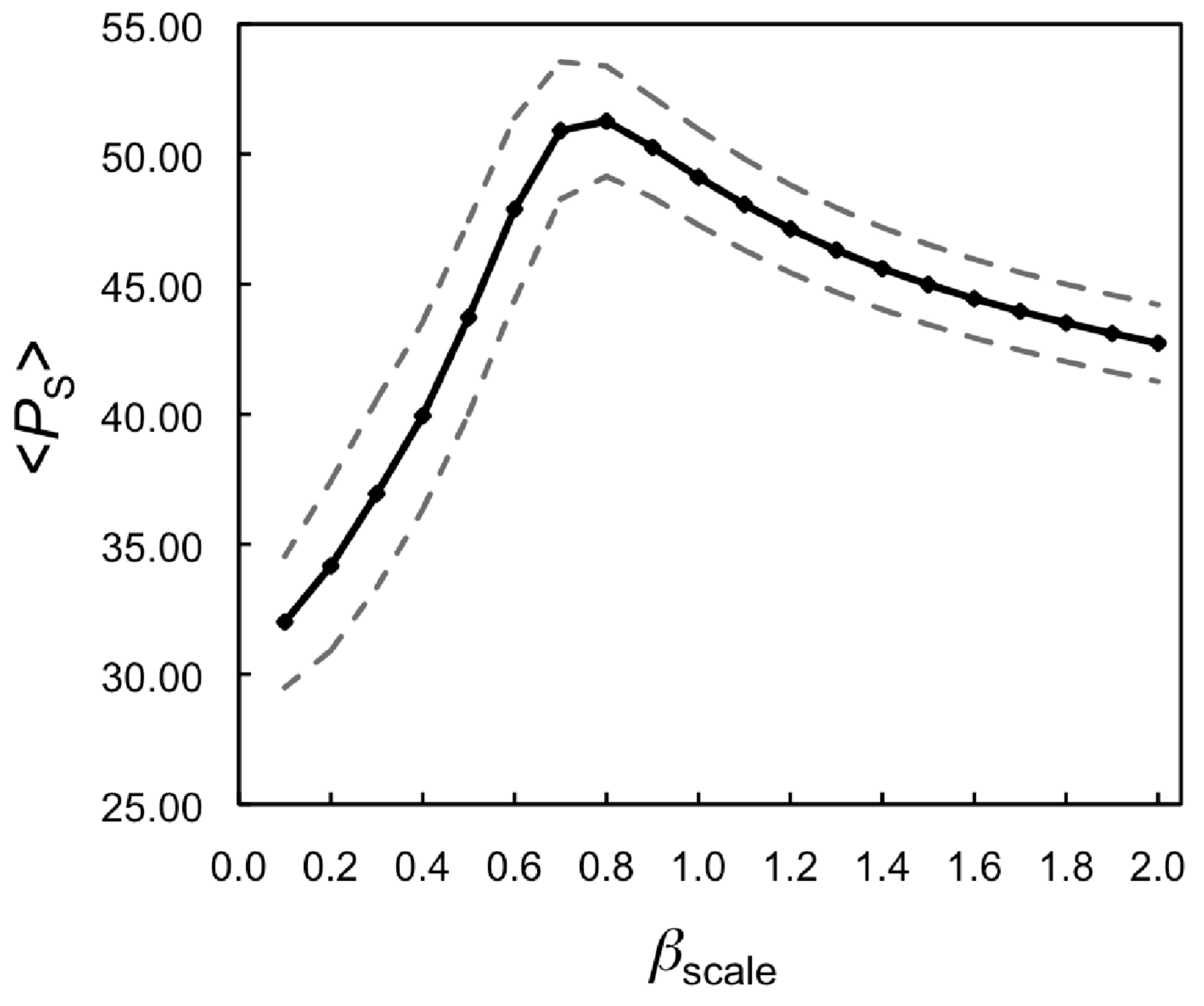}
\end{center}
\caption{To select an appropriate $\beta_{\rm scale}$ for classification of halo
shapes, we calculate the average peak signal-to-noise ratio, $\langle P_{\rm S} \rangle$,
over 200 input haloes (markers; black solid line).  Dashed lines represent the one-standard 
deviation error range.  On the basis of this analysis, we choose $\beta_{\rm scale} = 0.8$,
which presents  a reasonable comprimise to a full optimisation process.
\label{fig:betascale}} 
\end{figure}

\begin{figure*}
\begin{center}
\includegraphics[width=7.0in,angle=0]{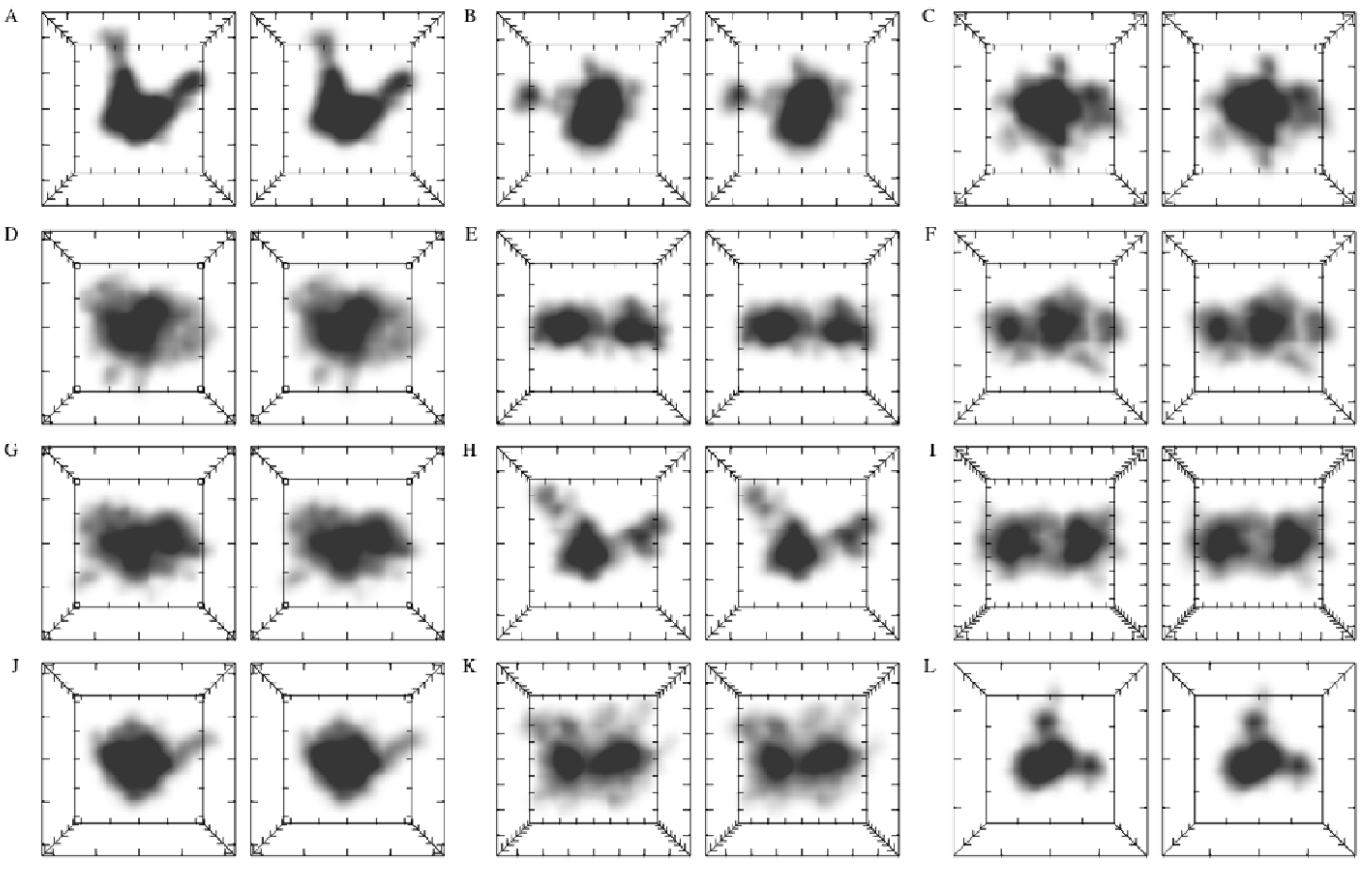}
\end{center}
\caption{The 12 most massive haloes from most massive (A; top left) 
to least massive (L; bottom right).  Each panel comprises (left) 
input dark matter halo and (right) shapelet reconstructed halo,
displayed as volume renderings of the logarithmic density.  
The coordinate ranges in each panel are not equal, but have been 
selected for clarity based on $\Delta x$ for each halo.
Shapelet parameters were $N_g = 51$, $n_{\rm max} = 24$, 
and $\beta$ values are in Table \ref{tbl:shapfits}.  
The strong similarity between the input and shapelet
reconstructed versions is apparent. 
\label{fig:heavy}} 
\end{figure*}

\begin{figure*}
\begin{center}
\includegraphics[width=7.0in,angle=0]{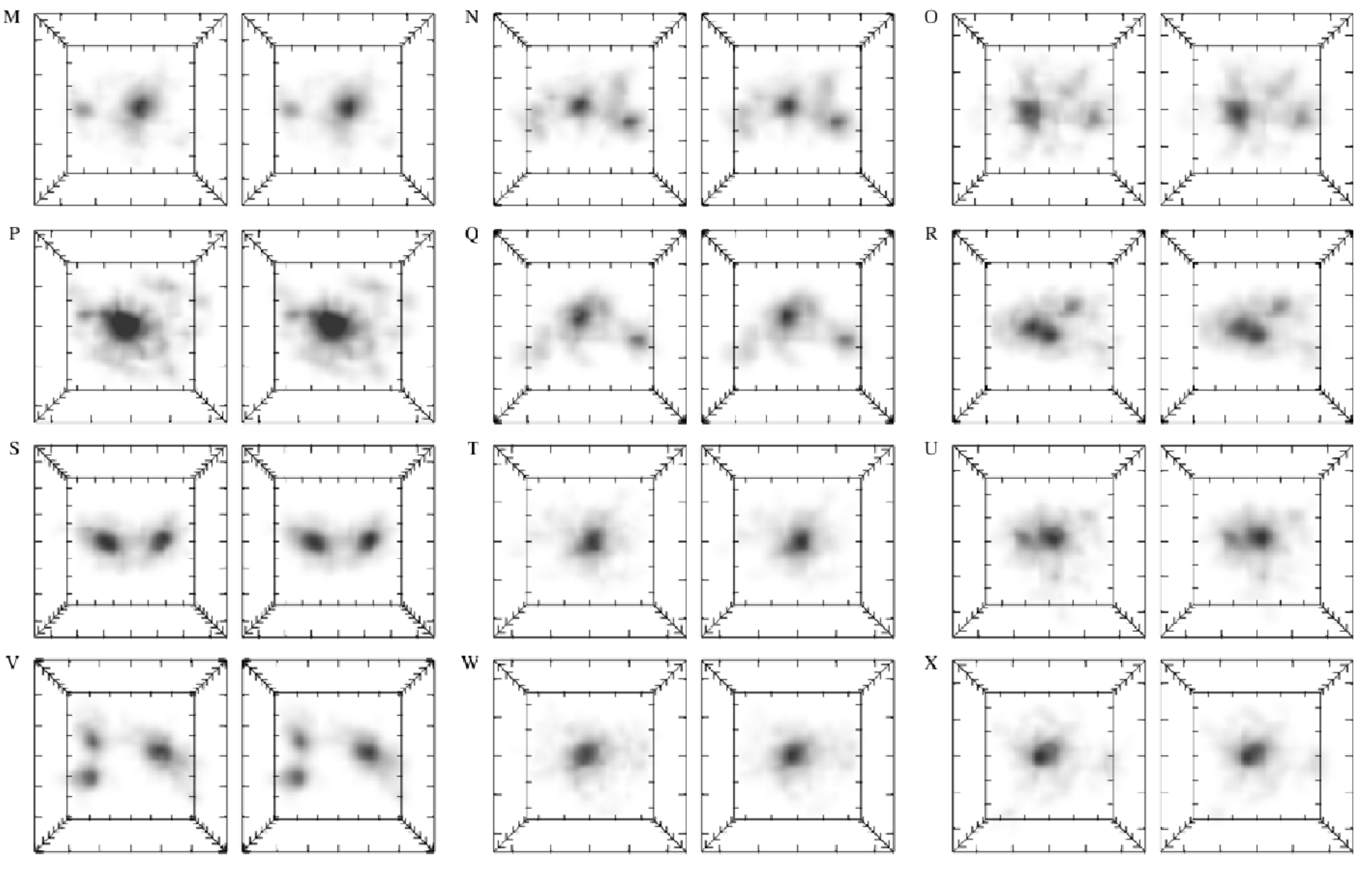}
\end{center}
\caption{The 12 least massive haloes from most massive (M; top left) 
to least massive (X; bottom right).  Each panel comprises (left) 
input dark matter halo and (right) shapelet reconstructed halo,
displayed as volume renderings of the logarithmic density.  
The coordinate ranges in each panel are not equal, but have been 
selected for clarity based on $\Delta x$ for each halo.
Shapelet parameters were $N_g = 51$, $n_{\rm max} = 24$, 
and $\beta$ values are in Table \ref{tbl:shapfits}.  
The strong similarity between the input and shapelet
reconstructed versions is apparent. 
\label{fig:light}} 
\end{figure*}

For the application domain we now explore, viz. 3-d Cartesian shapelet 
representations of simulated dark matter haloes, we have used a CPU-only 
implementation of the decomposition algorithm. 

\section{The shapes of dark matter haloes}
\label{sct:cosmoapp}
If the only use of the shapelet approach was to calculate the analytic 
expressions of Section 3, then it would be a somewhat ineffecient one, 
compared to direct numerical integration of 
equations (\ref{eqn:ana1}), (\ref{eqn:ana2}), 
(\ref{eqn:ana3}), (\ref{eqn:ana4}), and (\ref{eqn:ana5}).  The benefit of 
the shapelet decomposition is that we now have additional information 
concerning the shape. Opportunities for classifying three-dimensional 
structures based on the shapelet terms may be made through identification 
of the dominant shapelet terms, or by investigating relative weights 
of particular shapelet orders.  In this section, we demonstrate 
how three-dimensional shapelet analysis of dark matter halos suggests 
a new method for automatically classifying halo types. 

\subsection{Shapes and sub-structure}
\label{sec:haloes}
For some time, it has been known that Cold Dark Matter (CDM) cosmologies 
predict the formation of triaxial haloes (on average), with a 
slight preference for prolate haloes over oblate ones (Davis et al. 1985; 
Barnes \& Efstathiou 1987; Frenk et al. 1988; Dubinski \& Carlberg 1991; 
Dubinski 1994; Cole \& Lacey 1996; Jing \& Suto 2002; 
Kasun \& Evrard 2005; Bailin \& Steinmatz 2005; Oguri et al. 
2005; Allgood et al. 2006; Knebe \& Wie{\ss}ner 2006;
Kuhlen, Diemand \& Madau 2007).
These studies include measuring the distribution of halo triaxalities, 
studying the effects of baryons (which tend to reduce the 
triaxiality compared to dark matter only models),  and investigating 
the relationships between halo shapes and angular momentum.  

The purely triaxial treatment of dark matter haloes overlooks another 
well-established result from CDM simulations: individual haloes do not 
have a smooth density profile -- they contain sub-structure 
(Lacey \& Cole 1993; Moore et al. 1999; Ghigna et al. 2000). 
While the triaxial nature of dark matter haloes can be expressed 
empirically (e.g. Jing \& Suto 2002), quantifying the sub-structure 
remains a challenge.   A shapelet-space representation of
dark matter haloes provides a potential solution.

To demonstrate our approach, we use a sample of 200 candidate 
dark matter haloes selected from a cosmological $N$-body simulation 
performed with GADGET-2 (Springel 2005). The cosmological parameters 
were $\Omega_0 = 0.27$, $\Lambda_0 = 0.73$, $h = 0.71$ and
$\sigma_8 = 0.9$, and candidate haloes were identified  using the 
SubFind groupfinder (Springel et al. 2001).

Using particle number, $N_p$,  as a proxy for mass,
we pay particular attention to the twelve most massive
haloes, haloes A--L, and the twelve least massive, haloes M-X, from 
the sample.  We consider these two-subsets as being
representative of typical halo shapes and presence of sub-structure,
along with limiting any mass-dependent biases that may occur.   
For each halo, the triaxality, $T$, is calculated using 
the approach described in Appendix \ref{app:triax}, and tabulated
in Table \ref{tbl:shapfits}.  Further quantities presented in this 
table are described below.

Of the twelve `heavy' haloes, two are oblate ($T \leq 1/3$), eight are 
prolate ($T \geq 2/3)$ and two are triaxial ($1/3 < T < 2/3$). 
Both the oblate haloes (A and L) have clear central
cores, while the triaxial haloes (D and J) do not possess such a core. 
None of the `light' haloes are oblate, ten were prolate, 
and two were triaxial (this time, haloes with central cores).

\begin{table*}
\caption{Summary of halo properties and shapelet decomposition parameters
for the sample of 24 dark matter haloes, classified into heavy (A-L) and light (M-X) samples.  
Table columns are: halo identifier; number of particles in halo, $N_p$, a proxy for halo mass; 
halo triaxiality, $T$, determined from particle positions; scale parameter, $\beta$, used for
decomposition; the cell-width, $\Delta x$, as defined in equation (\ref{eqn:deltax});
the maximum voxel values from the input shape, $I_{\rm max}$, and the shapelet-recovered 
minimum and maximum voxel values, $S_{\rm min}$ and $S_{\rm max}$; and
$\Sigma_{I} = \sum_{i,j,k} f_{ijk}$ and $\Sigma_{S} = \sum_{i,j,k} \hat{f}_{ijk}$ are used to characterise
the recovered shapes, along with the peak signal-to-noise, $P_{\rm S}$.  
The $m$-th most dominant shapelet component has $\bmath{n} = \bmath{D}_{m}$, with amplitude 
$f_{3,\bmath{D}_m} = f_{m,{\rm max}}$, 
and $m = 1, 2, \dots$; except where indicated, $\bmath{D}_1 = (0,0,0)$, so we also present results for $\bmath{D}_2$.  Halo classes are assigned `by eye'
through inspection of the real-space, $C_I$, and shapelet-space, $C_S$, 
representations -- see Section  \ref{sct:auto} for details.
\label{tbl:shapfits}}
\begin{center}
\begin{tabular}{crcccccccccccccc}
Halo & $N_p$ & $T$ & $\beta$ & $\Delta x$ & $I_{\rm max}$ & $\Sigma_I$ & $S_{\rm min}$ & $S_{\rm max}$ & $\Sigma_S$ 
& $P_S$ & $f_{1,\rm max}$ & $\bmath{D}_1$ &  $\bmath{D}_{2}$ & $C_I$ & $C_S$ \\ 
\hline
A  & 1425030 & 0.239 & 0.85 & 0.42 & 4.64 & 2734.9 & -0.22 & 4.51 & 2739.8 & 47.27 & 15.307 && (2,0,0) & 2 & 2\\
B  & 62492 & 0.753 & 0.22 & 0.11 & 3.18 & 2089.5 & -0.07 & 3.11 & 2093.6 & 51.29 & 1.279 && (0,2,0) & 2 & 2\\
C  & 47535 & 0.707 & 0.18 & 0.09 & 3.17 & 2455.0 & -0.10 & 3.06 & 2457.2 & 50.27 & 0.900 && (2,0,0)& 2 & 2\\
D  & 45760 & 0.406 & 0.17 & 0.09 & 3.02 & 2245.5 & -0.07 & 2.94 & 2245.1 & 51.38 & 0.826 && (2,0,0)& 2 & 2\\
E  & 43091 & 0.973 & 0.24 & 0.12 & 2.88 & 1465.3 & -0.08 & 2.81 & 1465.8 & 49.73 & 0.848 && (2,0,0) & 3 & 3\\
F  & 39700 & 0.871 & 0.20 & 0.10 & 3.01 & 1715.0 & -0.07 & 2.91 & 1715.3 & 51.06 & 0.905 && (2,0,0)& 2 & 2-3 \\
G  & 37735 & 0.816 & 0.17 & 0.08 & 2.96 & 2132.9 & -0.08 & 2.89 & 2133.5 & 51.01 & 0.753 && (2,0,0)& 2 & 2-3 \\
H  & 35417 & 0.694 & 0.21 & 0.11 & 2.94 & 1406.5 & -0.08 & 2.86 & 1406.5 & 51.00 & 0.891 && (2,0,0)& 2 & 2 \\
I  & 28290 & 0.963 & 0.15 & 0.07 & 2.74 & 1934.3 & -0.06 & 2.62 & 1933.4 & 49.38 & 0.372 & (2,0,0) & (0,0,0) & 3 & 3\\
J  & 27189 & 0.659 & 0.18 & 0.09 & 2.64 & 1567.1 & -0.06 & 2.55 & 1568.0 & 51.98 & 0.744 && (2,0,0)& 2 & 2-3 \\
K  & 21336 & 0.766 & 0.12 & 0.06 & 2.53 & 1863.9 & -0.07 & 2.45 & 1858.4 & 47.46 & 0.313 && (2,0,0)& 3 & 3-2 \\
L  & 20476 & 0.144 & 0.16 & 0.08 & 2.63 & 1184.9 & -0.06 & 2.53 & 1185.3 & 54.14 & 0.569 && (2,0,0)& 2 & 2  \\
\hline
M  & 864 & 0.868 & 0.04 & 0.02 & 1.39 & 230.83 & -0.02 & 1.30 & 229.85 & 52.35 & 0.018 && (1,0,0)& 2 & 2-1 \\
N  & 855 & 0.934 & 0.05 & 0.02 & 1.24 & 252.54 & -0.02 & 1.15 & 252.41 & 51.01 & 0.016 && (2,0,0)& 2 & 2-1 \\
O  & 845 & 0.716 & 0.04 & 0.02 & 1.15 & 261.30 & -0.02 & 1.07 & 261.67 & 50.01 & 0.011 && (2,0,0)& 3 & 3-2 \\
P  & 834 & 0.760 & 0.04 & 0.02 & 0.77 & 271.24 & -0.02 & 0.75 & 270.65 & 45.70 & 0.014 && (2,0,0)& 2 & 1 \\ 
Q  & 829 & 0.818 & 0.05 & 0.02 & 1.21 & 237.36 & -0.03 & 1.14 & 236.67 & 50.82 & 0.017 && (0,1,0)& 2 & 2 \\
R  & 824 & 0.859 & 0.05 & 0.02 & 1.09 & 241.40 & -0.02 & 1.02 & 240.89 & 50.11 & 0.020 && (2,0,0)& 3 & 2 \\
S  & 821 & 0.984 & 0.06 & 0.03 & 1.14 & 212.54 & -0.02 & 1.09 & 212.77 & 52.70 & 0.024 & (2,0,0) & (4,0,0)& 3 & 3 \\
T  & 816 & 0.713 & 0.04 & 0.02 & 1.37 & 227.73 & -0.02 & 1.27 & 228.45 & 53.16 & 0.018 && (2,0,2)& 1 & 1 \\
U  & 797 & 0.422 & 0.04 & 0.02 & 1.14 & 229.69 & -0.02 & 1.07 & 229.43 & 51.36 & 0.018 && (2,0,0)& 2 & 2-1 \\
V  & 794 & 0.847 & 0.05 & 0.02 & 1.09 & 226.23 & -0.03 & 1.02 & 227.14 & 49.75 & 0.013 & (4,0,0)& (2,0,0)& 3 & 3 \\
W  & 788 & 0.876 & 0.04 & 0.02 & 1.32 & 228.40 & -0.02 & 1.25 & 227.70 & 52.35 & 0.017 && (1,0,0)& 1 & 1 \\
X  & 778 & 0.512 & 0.04 & 0.02 & 1.26 & 214.21 & -0.02 & 1.18 & 213.40 & 51.27 & 0.018 && (1,0,0)& 1 & 1 \\
\hline
\end{tabular}
\end{center}
\end{table*}

We perform a three-dimensional shapelet decomposition on each halo,
with the following input parameters fixed: $N_g = 51$, $n_{\rm max} = 24$
and $\beta = \beta_{\rm scale} x_{\rm max}/\sqrt{2 N_g}$.  
To select an appropriate $\beta_{\rm scale}$ for classification of halo
shapes, we define a fitness estimator in terms of the peak signal-to-noise ratio:
\begin{equation}
P_s = 20 \log_{10} \left[ \frac{\mbox{Max}(f_{ijk})}{\sqrt{M_s}}\right],
\label{eqn:ps}
\end{equation}
where $\mbox{Max}(f_{ijk})$ is the maximum value in the volume, 
and the mean-square error is:
\begin{equation}
M_s = \frac{1}{N_{\rm g}^3} \sum_{i,j,k=1}^{N_{\rm g}}  \vert
f_{ijk} - \hat{f}_{ijk} \vert^2.
\end{equation}
We calculate the average peak signal-to-noise ratio, $\langle P_{\rm S} \rangle$,
over 200 input haloes (Figure \ref{fig:betascale} -- markers; black solid line); dashed lines represent 
the one-standard deviation error range.  On the basis of this analysis, we choose $\beta_{\rm scale} = 0.8$
as providing the best fit to the input halo shapes, presenting  a reasonable comprimise to a full 
optimisation process.

To avoid orientation-dependent effects, haloes are rotated 
such that their principle axes are aligned with the coordinate axes (see 
Appendix \ref{app:triax}).  Halo particles are then smoothed to a grid  
using the triangle-shaped cloud smoothing strategy, providing number
counts per voxel, which is equivalent to a density, $\rho_{ijk}$.  
To deal with the large dynamic
range in $\rho_{ijk}$, the input shape is actually:
\begin{equation}
f_{ijk} = \log_{10}\left(1+ \rho_{ijk}\right).
\end{equation}

Since each halo has a different mass and hence physical extent, 
the $\beta$ value for each halo is different 
-- see Table \ref{tbl:shapfits}.  The other columns in this table are:
the cell-width, $\Delta x$, as defined in equation (\ref{eqn:deltax});
the maximum voxel value from the input shape, $I_{\rm max}$, and 
the minimum and maximum shapelet-recovered values, $S_{\rm min}$ 
and $S_{\rm max}$, respectively.   To enable quantitative  
comparisons between the input and reconstructed shapes 
we compute the quantitites $\Sigma_{I} = \sum_{i,j,k} f_{ijk}$ and $\Sigma_{S} 
= \sum_{i,j,k} \hat{f}_{ijk}$, and $P_s$.
Numerical testing, where reconstructions were optimised by hand, 
suggested that $\Sigma_{I} \sim \Sigma_S$ and $P_s \geq 45$ (Figure \ref{fig:betascale})
represented a good shapelet fit for the grid resolution used.  

The $m$-th most dominant shapelet component 
of the reconstruction has $\bmath{n} = \bmath{D}_{m}$, with amplitude 
$f_{3,\bmath{D}_m} = f_{m,{\rm max}}$.  
Except where indicated, $\bmath{D}_1 = (0,0,0)$, so we also report the 
value of $\bmath{D}_2$.
The final two columns of Table \ref{tbl:shapfits} represent the result of
`by-eye' classifications of the spatial characteristics of each halo,
$C(I)$, and the shapelet profiles, $C(S)$, into the three halo
classes -- see Section \ref{sct:auto} below.

Figs.~\ref{fig:heavy} and
\ref{fig:light} show the results of the shapelet decomposition.  For each
halo, the left-hand panel shows the input shape, and the right-hand 
panel is reconstructed in shapelet space.  Each image pair presents
two-dimensional projections of fully three-dimensional, volume rendered
structures.  
Visual comparsion of pairs of images suggests that, qualitatively, 
Cartesian shapelets represent an appropriate basis set for decomposition 
of dark matter haloes.   Quantitatively, we find that:
\begin{equation}
\left| \left(S_{\rm max} - S_{\rm min}\right) / I_{\rm max} - 1\right| \leq 6 \%,
\end{equation}
\begin{equation}
\left| \Sigma_S/ \Sigma_I  - 1\right| \leq 1 \%,
\end{equation}
and $P_S \geq 45$,
so that even without a halo-specific optimisiation, there is excellent
agreement between the input halo and its Cartesian shapelet reconstruction.

\begin{figure*}
\begin{center}
\includegraphics[width=7.0in,angle=0]{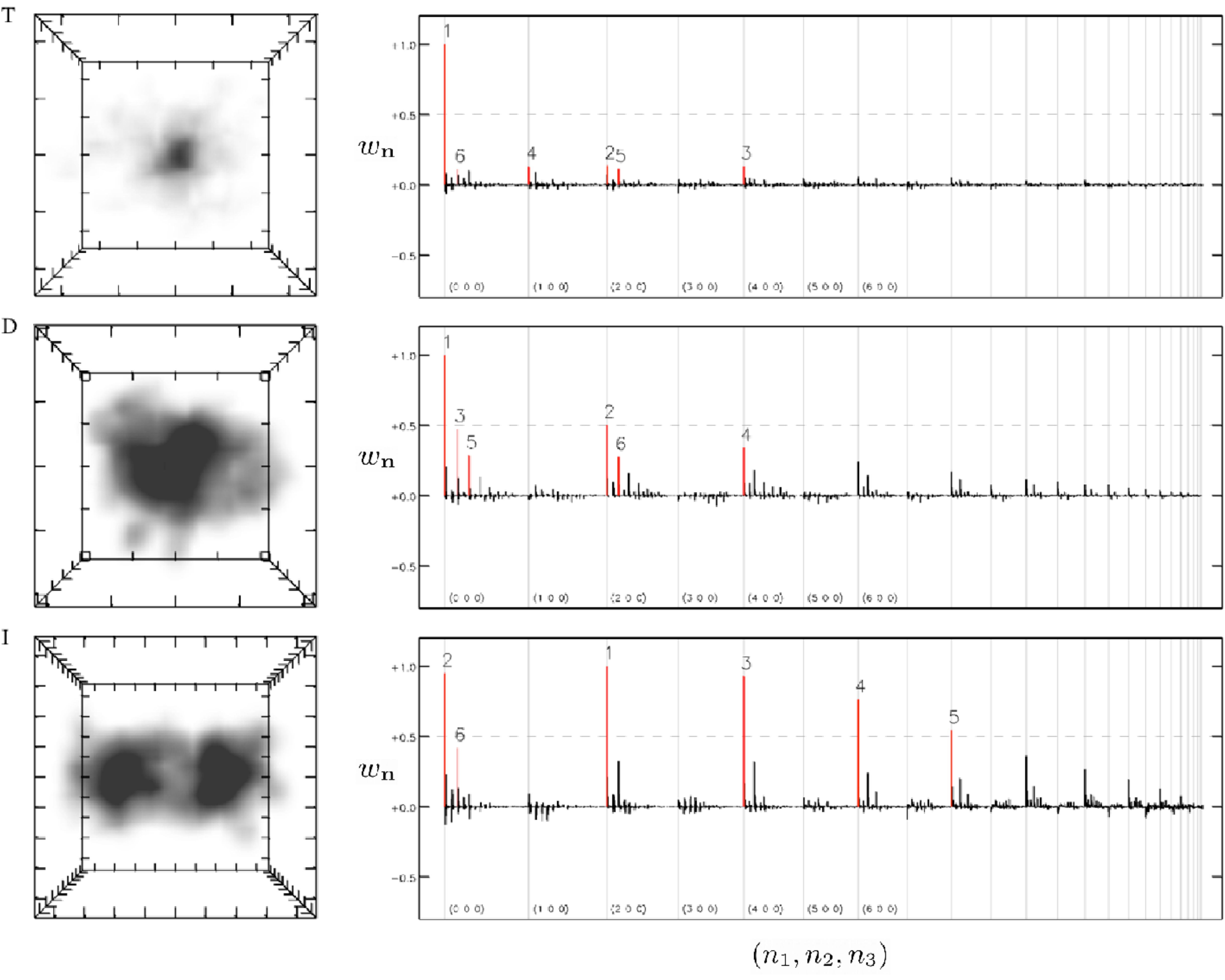}
\end{center}
\caption{Three characteristic shapelet-space representations of dark matter haloes.
Shapelet coefficient amplitudes are plotted in index order, with $n_3$ value varying most 
rapidly, then $n_2$, and finally $n_1$ -- the light grey vertical lines indicate 
values of $(n_1, 0, 0)$ with $n_1 = 0, 1, \dots, n_{\rm max}$.  Shapelet amplitudes, represented by vertical 
black line segments, are $w_{3,\bmath{n}} = f_{3,\bmath{n}}/f_{1,\rm max}$, with the six 
most dominant shapelet orders numbered and coloured red. 
(Top) Halo T, Class 1 -- central core, no significant sub-structure -- dominated
by zeroth-order shapelet, low-amplitude for higher orders.
(Middle) Halo D, Class 2 -- central core, significant sub-structure -- $f_{\rm max}$ occurs at $\bmath{n} = (0,0,0)$ and several
higher order shapelets have amplitudes $\sim \frac{1}{2} f_{\rm max}$.
Bottom) Halo I, Class 3 -- significant sub-structure, no central core 
-- zeroth-order 
shapelet is not the dominant term (although in other Class 3 haloes, it
can still be dominant), several orders with amplitudes $\sim f_{\rm max}$.
\label{fig:spikes} }
\end{figure*}

\begin{figure*}
\begin{center}
\includegraphics[width=7.0in,angle=0]{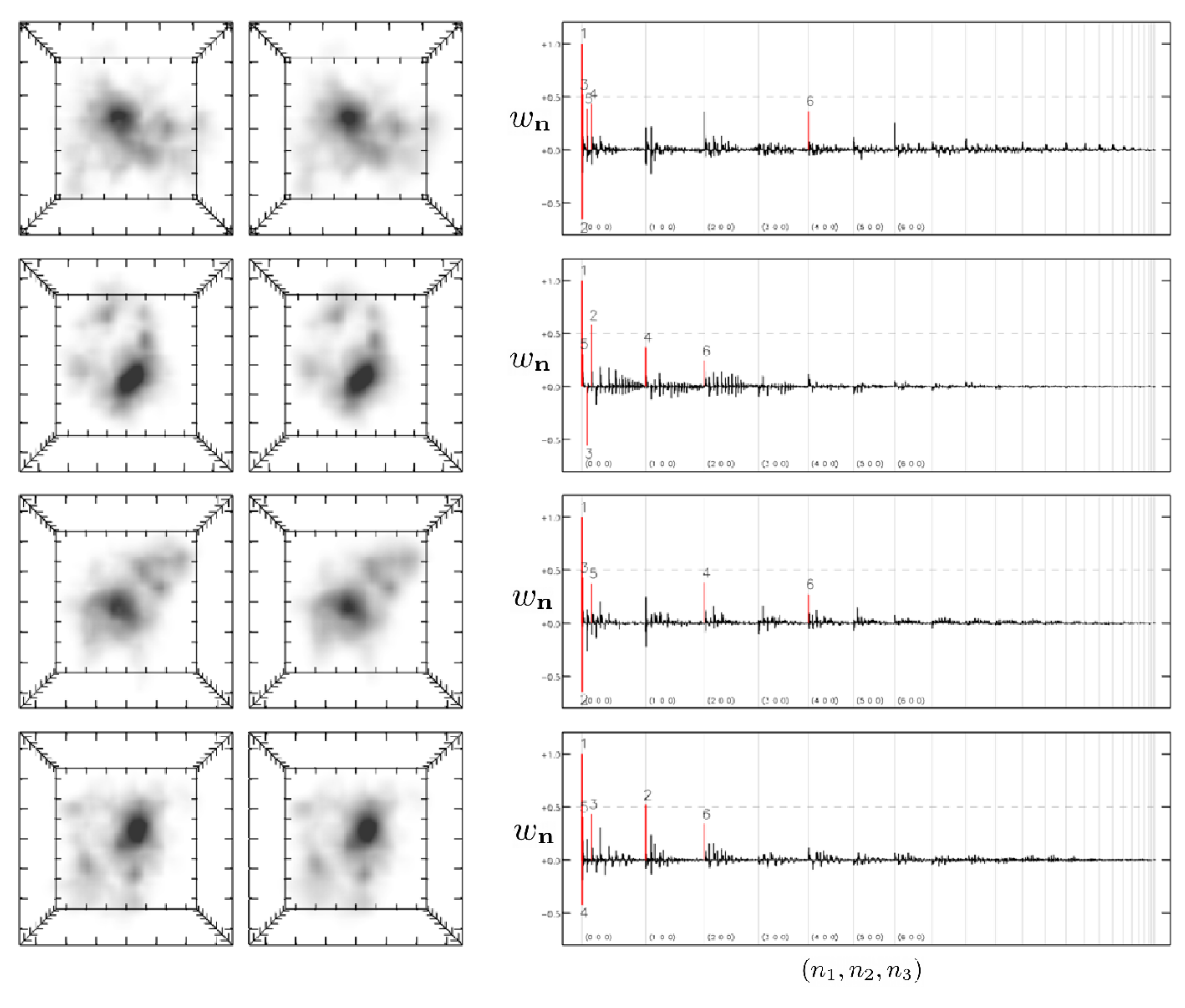}
\end{center}
\caption{Halo 100, $N_p = 2068$, $T = 0.80$ (prolate), Class 2.  The left-hand column shows four real-space
configurations of the halo, with arbitrary rotations about the centre-of-mass. The right-hand
column shows the corresponding shapelet-space configuration; $w_{\bmath{n}} = f_{3, \bmath{n}}/f_{1, {\rm max}}$,
and the horizonal axis represents the sequential coefficients, $\bmath{n}$ -- see Section \ref{sct:auto}.
General properties of the
distribution of shapelet amplitudes are preserved regardless of orientation.  \label{fig:rot1}}
\end{figure*}

\begin{figure*}
\begin{center}
\includegraphics[width=7.0in,angle=0]{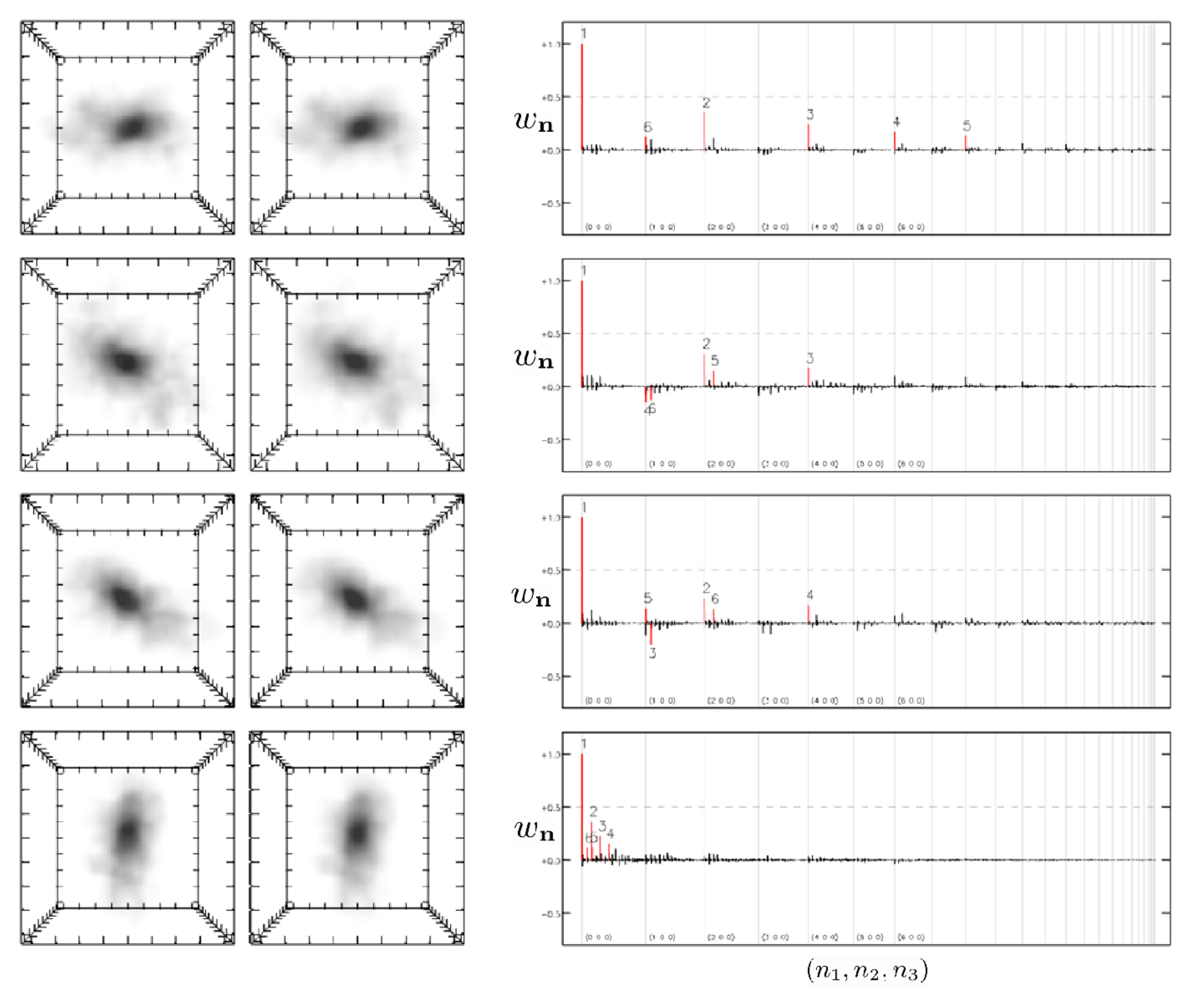}
\end{center}
\caption{Halo 101, $N_p = 2047$, $T = 0.92$ (prolate), Class 1. The left-hand column shows four real-space
configurations of the halo, with arbitrary rotations about the centre-of-mass. The right-hand
column shows the corresponding shapelet-space configuration; $w_{\bmath{n}} = f_{3, \bmath{n}}/f_{1, {\rm max}}$,
and the horizonal axis represents the sequential coefficients, $\bmath{n}$ -- see Section \ref{sct:auto}.
General properties of the
distribution of shapelet amplitudes are preserved regardless of orientation.  \label{fig:rot2}}
\end{figure*}

\begin{figure*}
\begin{center}
\includegraphics[width=7.0in,angle=0]{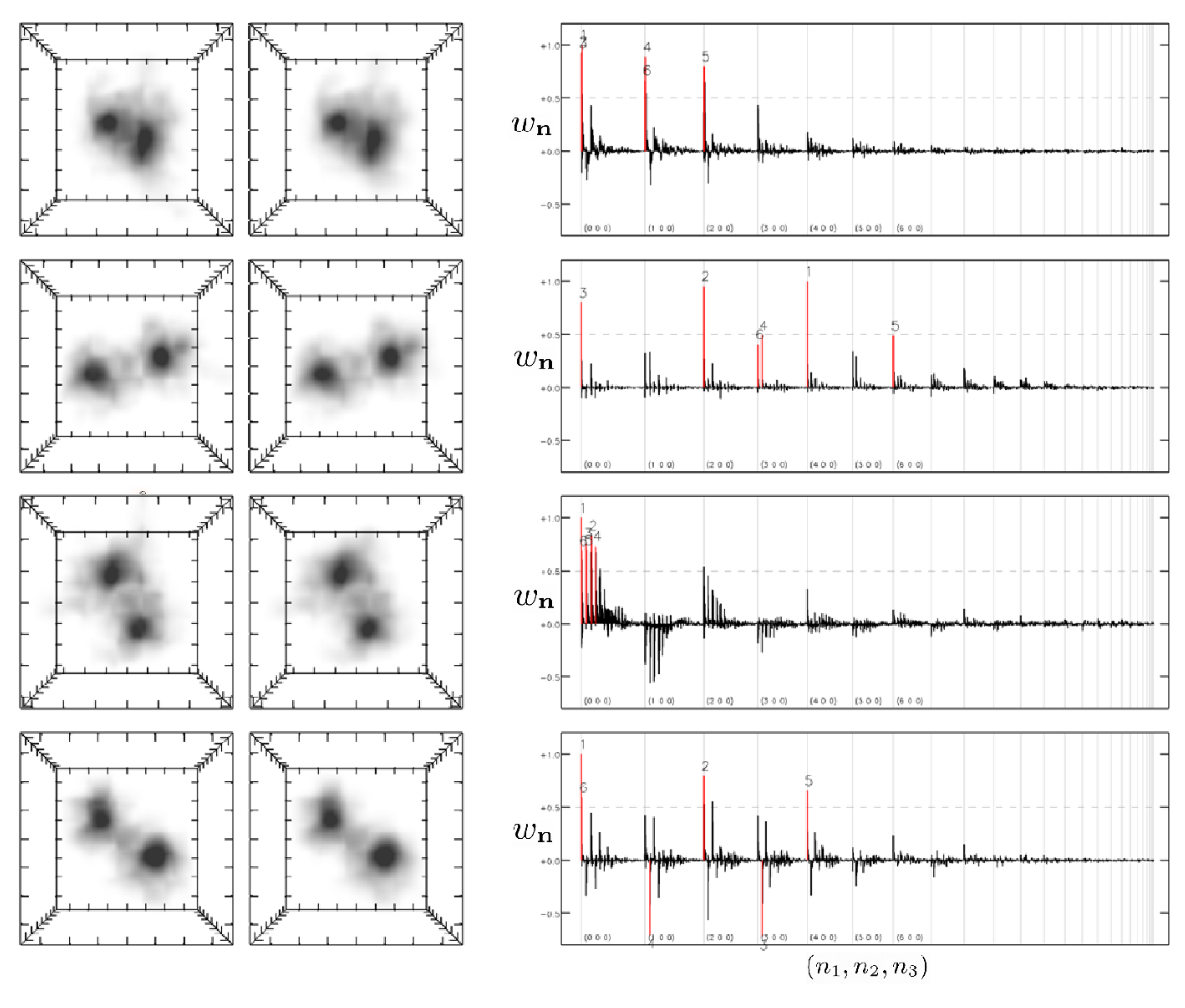}
\end{center}
\caption{Halo 102, $N_p = 2005$, $T = 0.91$ (prolate), Class 3.  The left-hand column shows four real-space
configurations of the halo, with arbitrary rotations about the centre-of-mass; the right-hand
column shows the corresponding shapelet-space configuration; $w_{\bmath{n}} = f_{3, \bmath{n}}/f_{1, {\rm max}}$,
and the horizonal axis represents the sequential coefficients, $\bmath{n}$ -- see Section \ref{sct:auto}.
General properties of the
distribution of shapelet amplitudes are preserved regardless of orientation.  \label{fig:rot3}}
\end{figure*}

\begin{figure*}
\begin{center}
\includegraphics[width=7.0in]{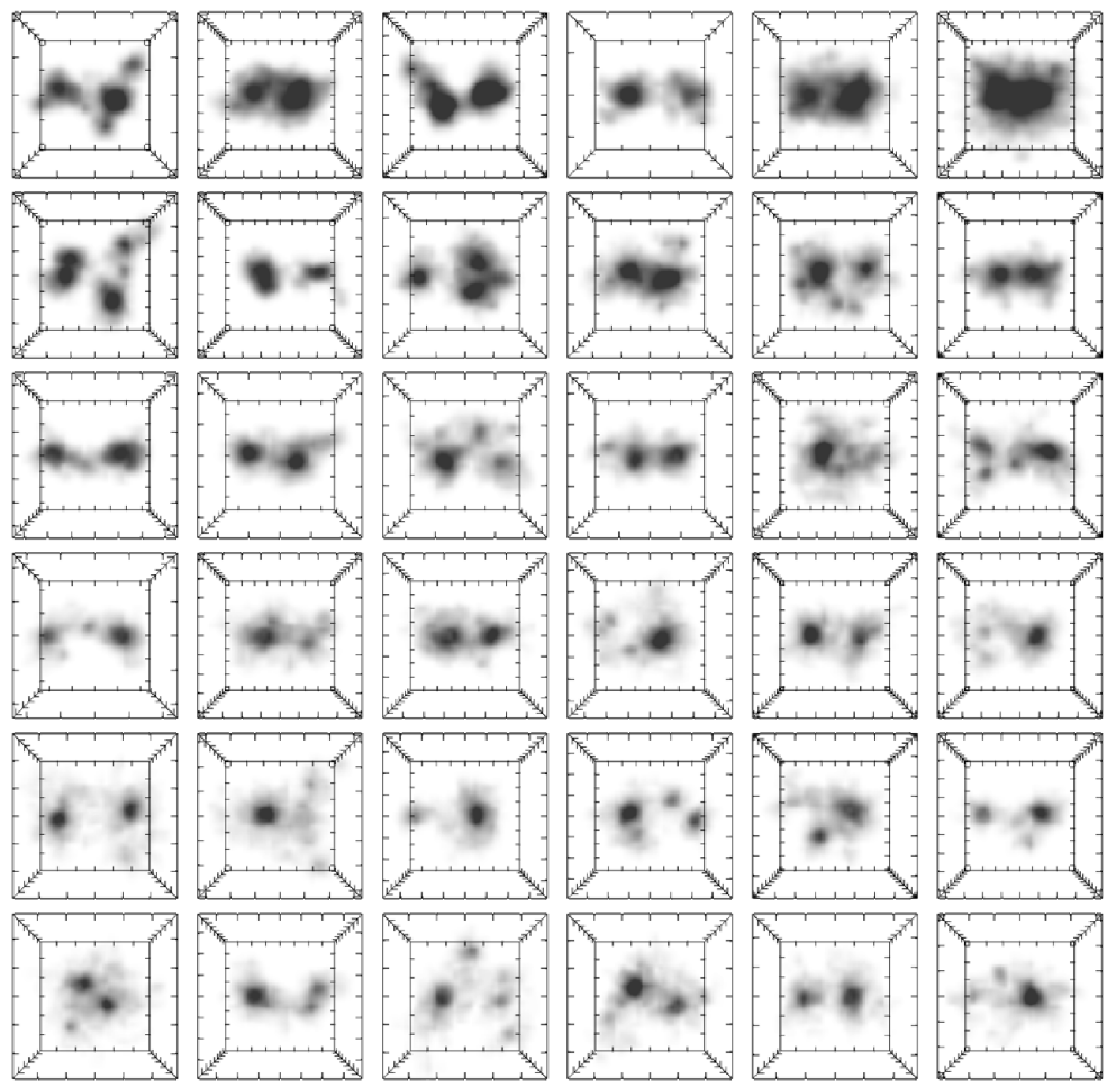}
\end{center}
\caption{From a visual investigation of 176 haloes in shapelet coeffecient 
space, 44 were selected as having obvious sub-structure and no central core.
36 of these haloes are shown here.  Visual
inspection of the remaining 130 haloes in real space suggests that 
an additional 10 should have been identified as Class 3. Further 
inspection in shapelet space confirmed the limitation of a `by-eye' classifier. 
\label{fig:plot36}}
\end{figure*}

\subsection{Towards an automated shape classifer}
\label{sct:auto}
The 3-d shapelet approach provides a means to check the outcome of 
halo finding algorithms by identifying classes in shapelet space 
without needing to visually inspect entire halo-candidate catalogues.
For 21 of the 24 haloes in Table \ref{tbl:shapfits}, 
the dominant component is $\bmath{D}_1 = (0,0,0)$ -- in most cases, the zeroth-order shape
has a high amplitude, which is not unexpected for haloes centred on the coordinate origin. 
Three haloes (I, S and V), however, receive their maximal contribution from a higher-order shapelet, 
${\bmath D}_1 = (n_1, 0, 0), n_1 \geq 1$.  We can use information on
the relative contributions of shapelet orders higher than the zeroth 
order term to enable a shapelet-based classification of dark matter 
halo shapes.

Fig.~\ref{fig:spikes} shows three characteristic patterns in shapelet
space, consistent with the general appearance of the haloes in Figs.~\ref{fig:heavy} and
\ref{fig:light}.  For each halo, all amplitudes are plotted in index order,
with $n_3$ value varying most rapidly, then $n_2$, and finally $n_1$.
The light grey vertical lines indicate values of 
$(n_1, 0, 0)$ where $n_1 = 0, 1, \dots, n_{\rm max}$; 
for $n_{\rm max} = 24$, there are $N_{\rm eval} = 2925$ shapelet coefficients.
Shapelet amplitudes, represented by vertical black line segments, are plotted 
as $w_{\bmath{n}} = f_{3,\bmath{n}}/f_{1,{\rm max}}$, with the six most-dominant shapelet orders numbered and coloured red. 
We propose the following three classes:
\begin{itemize} 
\item {\bf Class 1}: Halo T (top panel) has a central core, but no significant sub-structure.
In shapelet space, it is dominated by the zeroth-order shapelet, with low amplitudes for higher orders.
\item {\bf Class 2}: Halo D (middle panel) has a central core, and obvious sub-structure. Here, 
the zeroth-order shapelet again dominates, but there are several
higher order shapelets with amplitudes $\lesssim \frac{1}{2} f_{1,{\rm max}}$. 
\item {\bf Class 3}: Halo I (bottom panel) has significant sub-structure
and no central core. The zeroth shapelet is no longer always the 
dominant term, 
and there are several shapelet orders with amplitudes $\sim f_{1,{\rm max}}$.
\end{itemize}
The initial alignment of each halo with the $x$-axis is apparent, with 
obvious contributions from shapelet orders $\bmath{n} = (n_1, 0, 0)$.

The flexibility of the classification system is demonstrated in 
Figs.~\ref{fig:rot1}-\ref{fig:rot3}.  We select three new intermediate mass haloes:
Haloes 100, 101 and 102 (specific properties are listed in the captions). 
Performing shapelet decomposition on these haloes with
arbitrary three-dimensional rotations, thus removing the alignment of the
principle moments of inertia with the coordinate axes, we see that the basic
features of the three shapelet classes remain.  Haloes 100 and 101, with $f_{1,{\rm max}}$ 
occuring for the zeroth-order shapelet, retain this behaviour, while the power in higher
shapelet orders is distributed away from the $(n_1, 0, 0)$ values.  This is not 
unexpected from the behaviour of 2-d shapelets under rotations (see Refregier 2003). 
Rotation of Halo 102 (Fig.~\ref{fig:rot3}), with two clear components, results in 
variation in the highest-amplitude shapelet coeffecients, suggesting the following
features for identification of haloes of this type: either $f_{1,{\rm max}}$ occurs
for a shaplet order other than the zeroth-order, or there are one or more shapelet
orders with amplitudes $\gtrsim 0.5 f_{1,{\rm max}}$. 

We use this heuristic to now attempt a purely (by-eye) shapelet-based selection of 
haloes with clear multiple sub-structures (Class 3).  We apply the shapelet decomposition with the same 
input parameters as used throughout this initial implementation, to a total of 176 haloes.
Particle counts for this new set of haloes are in the range $865 \leq N_p \leq 20033$, 
noting that these haloes are at intermediate masses to the 24 investigated previously.  
We identify 44 Class 3 haloes on the basis of their shapelet representation, the first 36 of which
are shown in Fig.~\ref{fig:plot36}, and all of which exhibit the expected spatial characteristics. 
Visual inspection of the remaining 132 haloes suggests a futher 10 haloes that should have
been identified from their shapelet representations.  In all cases, reinvestigation
of the shapelet distribution revealed that they were very close to meeting the 
criteria for a Class 3-halo.  While a more robust approach to classification 
is required for a full implementation (e.g. using an appropriately-sized training 
set and the construction of a decision tree or neural network classifier), our
results do suggest that there is benefit to performing classification 
of dark matter halo shapes in shapelet space.  

The existence of multiple cores in the dark matter haloes has implications for
computation of halo triaxiality -- the classification of `heavy' haloes E, F, I, K 
as prolate based purely on the principle moments of inertia is somewhat misleading;
in each case, an argument could be made that an isolated group has not been 
identified using, in this case, the Subfind algorithm.  Our application to 176 haloes 
identifies 44(+10) prolate haloes where the inferred triaxality was based on counting
potentially distinguishable sub-haloes as a single halo.  A shapelet-based automated 
classifier provides a method of identifying such haloes without needing to 
visually inspect each halo.  

\begin{figure}
\begin{center}
\includegraphics[width=3.4in,angle=0]{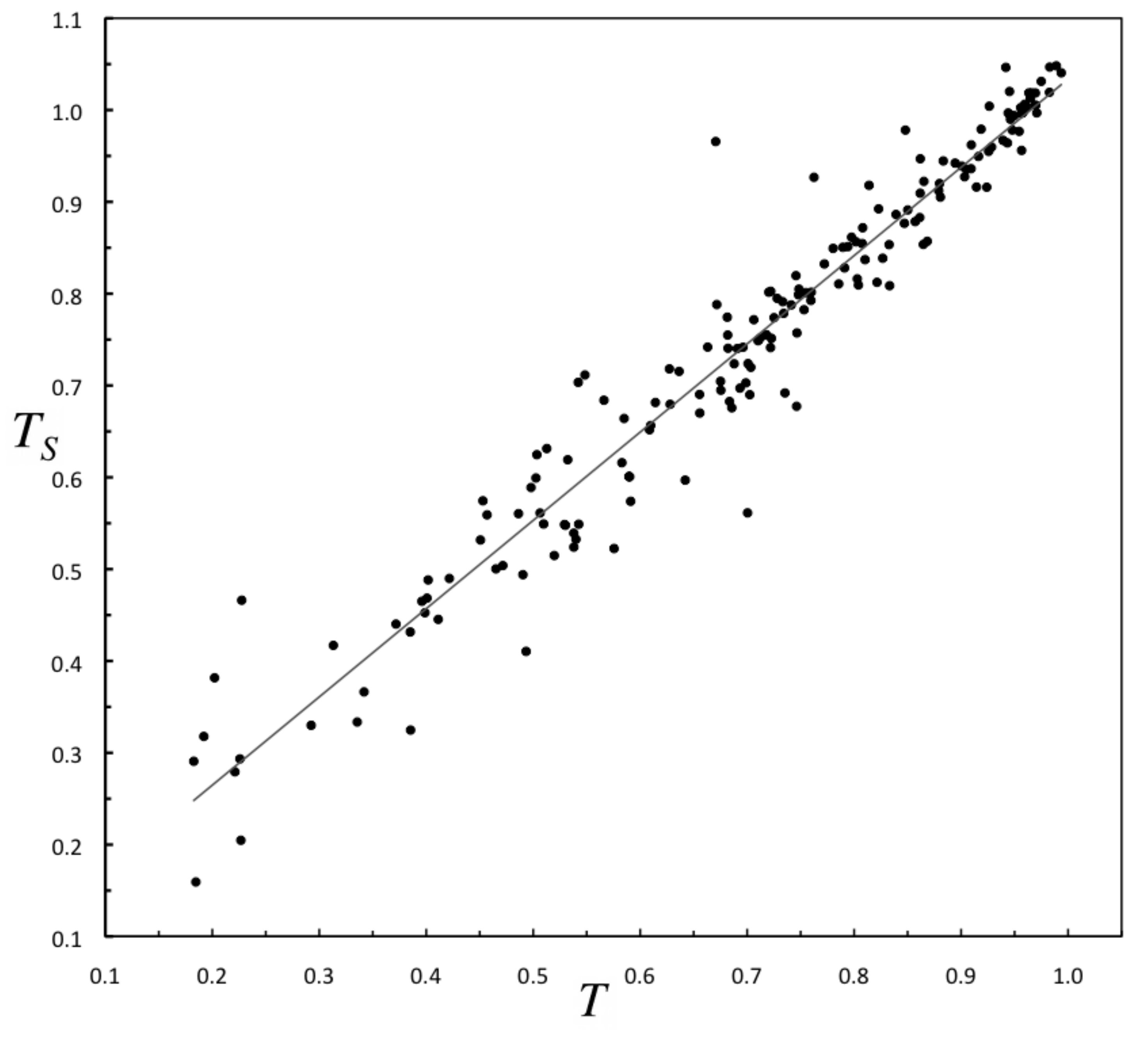}
\end{center}
\caption{The shapelet-based triaxiality, $T_S$, plotted against the
value determined from the original particle positions, $T$, for the sample of 176 
intermediate mass haloes.  A least-squares fit to the data (solid line) gives 
$T_S = 0.96 T + 0.02$ with the Pearson coeffecient, $r = 0.98$.
\label{fig:TvsT}}
\end{figure}

\begin{figure}
\begin{center}
\includegraphics[width=3.4in,angle=0]{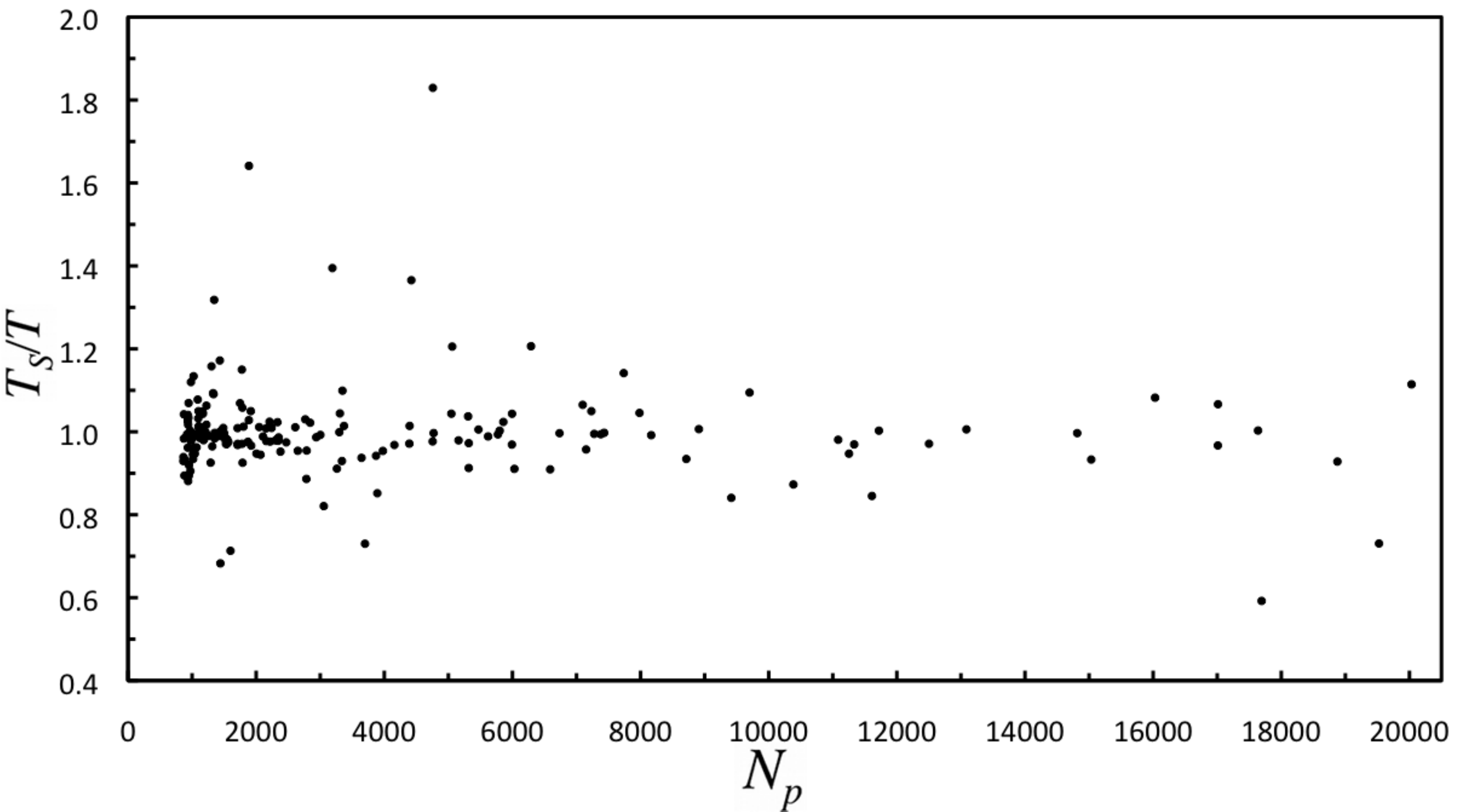}
\end{center}
\caption{The ratio of the shapelet-based triaxiality to the particle-based 
value, $T_S/T$, plotted against the number of particles, $N_p$, in each 
of 176 intermediate mass haloes.  We find that $\langle T_s/T \rangle =
0.99 \pm 0.12$, where the error is the sample standard deviation.
\label{fig:TvsN}}
\end{figure}

\subsection{Shapelet-based quantification}
\label{sct:quant}
As an example of quantitative analysis in shapelet space, we calculate the moment of
inertia tensors from equation (\ref{eqn:ana6})--(\ref{eqn:ana7}) and hence triaxiality, $T$.  
In Fig.~\ref{fig:TvsT}, we plot the shapelet-based triaxiality, $T_S$, against the
value determined from the original particle positions, $T$, for the sample of 176 
intermediate mass haloes.  A least-squares fit to the data (solid line) gives 
$T_S = 0.96 T + 0.02$ with the Pearson coeffecient, $r = 0.98$.
In Fig.~\ref{fig:TvsN}, we plot the ratio of $T_S/T$ against the 
particle number, $N_p$, which suggests that there is a slightly larger scatter for the
lower mass haloes.  We find that $\langle T_s/T \rangle = 0.99 \pm 0.12$, where the
error is the sample standard deviation.  Even though we have not performed a
per-halo optimisation for ($\beta$, $n_{\rm max}$, $\bmath{x}_c$), the shapelet-based
analytic result does indeed provide a very good estimator for the halo triaxiality.

\section{Summary and outlook}
\label{sct:conc}
We have extended the two-dimensional Cartesian shapelet formalism 
of Refregier (2003) to three dimensions, deriving analytic expressions
for the zeroth moment, object centroid, root-mean-square radius, 
and the components of the quadrupole moment and moment of inertia tensors.  
We also presented generalisations to $d$-dimensions. 

Further work is necessary to develop a robust and systematic optimisation 
strategy for the decomposition parameters, and the development of
specfic applications for the three-dimensional shapelet technique 
requires such a strategy. There are also opportunities to develop 
the formalism further, specifically extending it 
to include spherical shapelet functions [c.f. the alternative 
presentation of two-dimensional Cartesian shapelets as polar shapelets
by Massey \& Refregier (2005)].

The shapelet decomposition algorithm exhibits attributes that make it
an ideal target for implementation on modern, massively-parallel GPUs.
Our algorithm analysis demonstrates that the computation is entirely
(or embarassingly) parallel; has minimal or no branching; maintains a 
high ratio of arithmetic operations to memory-access operations; and 
has a memory access pattern that will result in aligned or contiguous
access to memory, required for achieving a high memory throughput.
With our proposed scheme of precomputing shapelet voxel-integral terms,
the computation reduces to a parallel series of multiply-add operations,
which are almost ideal for GPUs -- we anticipate achieving close to peak
processing performance.  Significantly reducing the computation time for
the shapelet  decomposition, compared to CPU, means that more processing time 
is then available for optimisation.

As an example application, we have demonstrated how three-dimensional 
shapelets can be used to study the complex sub-structures of dark matter haloes
from cosmological $N$-body simulations, including providing an alternative approach 
to classifying the properties of haloes.  Our preliminary investigation 
suggests that halo triaxiality measured purely from the moment of inertia 
tensor may be incorrect due to limitations of group finders that are not 
able to separate out what may be truly distinct sub-clumps.  Improvements to 
our current `by eye' approach to classification could include development of  
a decision tree or neural network classifier, or the use of principle
component analysis to significantly reduce the number of shapelet terms required
for classification (Kelly \& McKay 2004).

The shapelet formalism is virtually unexplored in the three-dimensional 
domain,  offering opportunities for the further development of a methodology 
that can be used to quantify and analyse complex three-dimensional
structures.  Future applications of the three-dimensional shapelet techinique
may include classification and parameterisation of sources identified in H{\sc i} 
spectral line data cubes; studying the shapes of voids in cosmological simulations
(by considering an inverted density field); and the possibility to generate
mock dark matter haloes through an extensive study of the distribution of shapelet
amplitudes as a function of mass and triaxiality.

\section*{Acknowledgments}
This research was supported under the Australian Research Council's 
Discovery Projects funding scheme (project number DP0665574).  
PL is supported by the Alexander von Humboldt Foundation.
CJF is grateful to Michael Vanner and Toffa Beer for their 
contributions to this work.  We thank Chris Power for providing the 
dark matter halo sample, David Bacon for early discussions on shapelets,
and our referee for his insightful comments.  Three-dimensional visualisation 
was conducted with the S2PLOT progamming library (Barnes et al. 2006).


\appendix
\section{Hermite polynomials}
\label{app:hermite}
We collect here a number of key expressions relating to Hermite polynomials,
which prove useful in deriving the analytic properties of Section 
\ref{sct:analytic}.

Expressing the Hermite polynomials via the Rodrigues formula
\begin{equation}
H_{n}(x)=(-1)^{n}{\rm e}^{x^{2}}\frac{d^{n}}{dx^{n}}\left({\rm e}^{x^{2}}\right),
\label{eqn:rodrig}
\end{equation}
one can show the important recursion relation
\begin{equation}
H_{n+1}(x)=2xH_{n}(x)-\frac{dH_{n}(x)}{dx},
\end{equation}
which further implies the shaplet basis functions satisfy:
\begin{equation}
\sqrt{2(n+1)}\beta B_{n+1}(x; \beta)=
\left(x -\beta^{2}\frac{{\rm d} }{{\rm d} x} \right) B_{n}(x;\beta)
\label{eqn:Bn+1}
\end{equation}
and
\begin{equation}
\left( x^2 - \beta^{4} \frac{{\rm d}^2 }{{\rm d} x^2} \right)
B_n(x;\beta)
= (2n+1) \beta^2 B_{n}(x;\beta),
\label{eqn:eigen}
\end{equation}
which is the eigenvalue equation.  Calculating the derivative terms, we have
the further recurrence relations:
\begin{equation}
H_{n+1}(x) = 2x H_{n}(x) - 2n H_{n-1}(x).
\end{equation}

\section{Triaxiality and halo rotations}
\label{app:triax}
Triaxiality of a dark matter halo is most easily expressed in terms of
the principle moments of inertia.  The principle moments, $(I_1, I_2, I_3)$, 
and the associated principle axes, $(\bmath{e}_1, \bmath{e}_2, \bmath{e}_3)$,
are the eigenvalues and eigenvectors of the moment of inertia tensor,
$\hat{I}$, respectively.
Approximating an arbitrary halo as a triaxial ellipsoid of the form
\begin{equation}
\frac{x^2}{a^2} + \frac{y^2}{b^2} + \frac{z^2}{c^2} = 1
\end{equation}
with $c \leq b \leq a$, then
\begin{eqnarray}
I_1 &=& \frac{M}{5}(b^2 + c^2)\\
I_2 &=& \frac{M}{5}(a^2 + c^2)\\
I_3 &=& \frac{M}{5}(a^2 + b^2)
\end{eqnarray}
and $I_1 \leq I_2 \leq I_3$. 
Moreover, we can calculate the triaxiality parameter (Franx, Illingworth \& de Zeeuw 1991):
\begin{equation}
T = \frac{a^2 - b^2}{a^2 - c^2} = \frac{I_2-I_1}{I_3-I_1},
\end{equation}
enabling us to classify haloes as oblate ($T \leq 1/3$), triaxial ($1/3 < T < 2/3$),
or prolate ($T \geq 2/3$).  We define a sphere ($a=b=c$) to have $T \equiv 0$.

Since we have full information on particle positions, $(x,y,z)$, from
the cosmological simulation, we make use of this to simplify the computation
of $\hat{I}$. Specifically, we compute elements of $\hat{I}$ from
particle positions, using standard eigensystem routines from the GNU Scientific 
Library\footnote{{\tt http://www.gnu.org/software/gsl/manual/html\_node/Eigensystems.html}}
to solve for the principle eigenvectors and eigenvalues. To enable comparisons
between halos, we rotate each halo so that its principle axes are aligned
with the Cartesian axes. First, we define an
orthogonal coordinate system with unit vector directions 
$\bmath{e}_1, \bmath{e}_2$ and 
\begin{equation}
\bmath{n} = \bmath{e}_1 \otimes \bmath{e}_2 .
\end{equation}
Although $\bmath{e}_3$ is orthogonal to $\bmath{e}_1$ and
$\bmath{e}_2$, explictly calculating the cross-product ensures that we have
a right-handed coordinate system.
The Euler angles are:
\begin{eqnarray}
\theta_1 &=& \sin^{-1} \left(-\bmath{e}_{2,z} \right)\\
\theta_2 &=& \tan^{-1} \left(\bmath{e}_{1,z}/\bmath{n}_z \right)\\
\theta_3 &=& \tan^{-1} \left(\bmath{e}_{2,x}/\bmath{2}_y \right).
\end{eqnarray}
We use the standard C-function {\tt atan2}, which returns the principle value
$\tan^{-1} (y/x)$, for calculating $\theta_2$ and $\theta_3$, and is
recommended for converting between rectangular and polar coordinates.

Next, we build a general $3 \times 3$ rotation matrix:
\begin{equation}
{\mathbfss R} = \left(
\begin{array}{ccc}
C_y C_z + S_x S_y S_z & C_x S_z & -S_y C_z + S_x C_y S_z \\
-C_y S_z + S_x S_y C_z & C_x C_z & S_y S_z + S_x C_y C_z \\
C_x S_y & -S_x & C_x C_y
\end{array}
\right),
\end{equation}
where $C_{x} = \cos(\theta_x)$, $S_{x} = \sin(\theta_y)$ and similarly
for $y$ and $z$, from which we can determine the 
inverse matrix, ${\textbfss R}^{-1}$, 
most easily via the transpose, ${\textbfss R}^T$, and the adjunct matrix of 
${\textbfss R}^{T}$.  Each particle position, $\bmath{p}$, in the halo is now rotated 
around the origin to new coordinates:
\begin{equation}
\bmath{p}' = {\textbfss R}^{-1} \bmath{p}.
\end{equation} 

\label{lastpage}

\end{document}